\documentclass[%
 reprint,
 superscriptaddress,
 showpacs,preprintnumbers,
 amsmath,amssymb,
 aps,
 prl,
]{revtex4-1}

\usepackage{changes}
\usepackage{graphicx}
\usepackage{dcolumn}
\usepackage{bm}
\usepackage{makecell}
\usepackage{braket} 
\usepackage{siunitx}
\usepackage{color}
\usepackage{xr}
\usepackage{upgreek}
\usepackage[colorlinks]{hyperref}
\hypersetup{colorlinks=true, citecolor=blue, urlcolor=blue, linkcolor=blue}
\usepackage[capitalize]{cleveref}
\usepackage[caption=false]{subfig}
\usepackage{multirow} 
\begin{document}
    \title{Long-Range $ZZ$ Interaction via Resonator-Induced Phase in Superconducting Qubits}
    
    \author{Xiang Deng}
	\thanks{These authors contributed equally to this work.}
    \affiliation{National Laboratory of Solid State Microstructures, School of Physics, Nanjing University, Nanjing 210093, China}
    \affiliation{Shishan Laboratory, Suzhou Campus of Nanjing University, Suzhou 215163, China}
    \author{Wen Zheng}
	\thanks{These authors contributed equally to this work.}
	\email{zhengwen@nju.edu.cn}
    \affiliation{National Laboratory of Solid State Microstructures, School of Physics, Nanjing University, Nanjing 210093, China}
    \affiliation{Shishan Laboratory, Suzhou Campus of Nanjing University, Suzhou 215163, China}
    \author{Xudong Liao}
    \affiliation{National Laboratory of Solid State Microstructures, School of Physics, Nanjing University, Nanjing 210093, China}
    \affiliation{Shishan Laboratory, Suzhou Campus of Nanjing University, Suzhou 215163, China}
    \author{Haoyu Zhou}
    \affiliation{National Laboratory of Solid State Microstructures, School of Physics, Nanjing University, Nanjing 210093, China}
    \affiliation{Shishan Laboratory, Suzhou Campus of Nanjing University, Suzhou 215163, China}
    \author{Yangyang Ge}
    \affiliation{National Laboratory of Solid State Microstructures, School of Physics, Nanjing University, Nanjing 210093, China}
    \affiliation{Shishan Laboratory, Suzhou Campus of Nanjing University, Suzhou 215163, China}

    \author{Jie Zhao}%
    \affiliation{National Laboratory of Solid State Microstructures, School of Physics, Nanjing University, Nanjing 210093, China}
    \affiliation{Shishan Laboratory, Suzhou Campus of Nanjing University, Suzhou 215163, China}
    \author{Dong Lan}%
    \affiliation{National Laboratory of Solid State Microstructures, School of Physics, Nanjing University, Nanjing 210093, China}
    \affiliation{Shishan Laboratory, Suzhou Campus of Nanjing University, Suzhou 215163, China}
    \affiliation{Synergetic Innovation Center of Quantum Information and Quantum Physics, University of Science and Technology of China, Hefei, Anhui 230026, China}
    \affiliation{Hefei National Laboratory, Hefei 230088, China}
    \author{Xinsheng Tan}%
    \affiliation{National Laboratory of Solid State Microstructures, School of Physics, Nanjing University, Nanjing 210093, China}
    \affiliation{Shishan Laboratory, Suzhou Campus of Nanjing University, Suzhou 215163, China}
    \affiliation{Synergetic Innovation Center of Quantum Information and Quantum Physics, University of Science and Technology of China, Hefei, Anhui 230026, China}
    \affiliation{Hefei National Laboratory, Hefei 230088, China}
    \author{Yu Zhang}
    \affiliation{National Laboratory of Solid State Microstructures, School of Physics, Nanjing University, Nanjing 210093, China}
    \affiliation{Shishan Laboratory, Suzhou Campus of Nanjing University, Suzhou 215163, China}
    \author{Shaoxiong Li}%
    \affiliation{National Laboratory of Solid State Microstructures, School of Physics, Nanjing University, Nanjing 210093, China}
    \affiliation{Shishan Laboratory, Suzhou Campus of Nanjing University, Suzhou 215163, China}
    \affiliation{Synergetic Innovation Center of Quantum Information and Quantum Physics, University of Science and Technology of China, Hefei, Anhui 230026, China}
    \affiliation{Hefei National Laboratory, Hefei 230088, China}
    \author{Yang Yu}%
    \email{yuyang@nju.edu.cn}
    \affiliation{National Laboratory of Solid State Microstructures, School of Physics, Nanjing University, Nanjing 210093, China}
    \affiliation{Shishan Laboratory, Suzhou Campus of Nanjing University, Suzhou 215163, China}
    \affiliation{Synergetic Innovation Center of Quantum Information and Quantum Physics, University of Science and Technology of China, Hefei, Anhui 230026, China}
    \affiliation{Hefei National Laboratory, Hefei 230088, China}

    \date{\today}

    \begin{abstract}
        {
    		Superconducting quantum computing emerges as one of leading candidates for achieving quantum advantage.
            However, a prevailing challenge is the coding overhead due to limited quantum connectivity, constrained by nearest-neighbor coupling among superconducting qubits.
            Here, we propose a novel multimode coupling scheme using three resonators driven by two microwaves, based on the resonator-induced phase gate, to extend the $ZZ$ interaction distance between qubits.
            We demonstrate a CZ gate fidelity exceeding 99.9\% within 160 ns at free spectral range (FSR) of 1.4 GHz, and by optimizing driving pulses, we further reduce the residual photon to nearly $10^{-3}$ within 100 ns at FSR of 0.2 GHz.
            These facilitate the long-range CZ gate over separations reaching sub-meters,
            thus significantly enhancing qubit connectivity and making a practical step towards the scalable integration and modularization of quantum processors.
            Specifically, our approach supports the implementation of quantum error correction codes requiring high connectivity, such as low-density parity check codes that paves the way to achieving fault-tolerant quantum computing.
        }
    \end{abstract}

    \maketitle

\textcolor{blue}{\textit{Introduction.}}---
Achieving fault-tolerant quantum computation is a necessary path to reach universal quantum computer \cite{nielsen_chuang_2010, shor_algorithms_1994, gottesman2010introduction, preskill2012, Georgescu2014, gambetta2017building}.
Conventional quantum-error-correcting codes (QECs) \cite{kitaev1997quantum, shor_faulttolerant_1997, kitaev_faulttolerant_2003, Fowler2012} suggest that millions of qubits is required, a target that currently overshoots the capabilities of existing quantum processors \cite{preskill_quantum_2018}.
To bridge this gap, advancements in connectivity are essential to lower the overheads associated with error correction. 
On the one hand, utilizing a modular architecture composed of multiple small high-yielding chips to form a large quantum processor is a promising approach to scale up physical qubits \cite{Jiang2007}.
In this approach, it is necessary to establish remote connections between chips to achieve deterministic entangling gates \cite{Xu2022}. 
On the other hand, since required entangling gates between non-local physical qubits \cite{Kim2013, cohen2022low, Tremblay2022, Zeng2019}, recently proposed quantum low-density parity check (LDPC) codes can significantly reduce the scale of physical qubits required to code logical qubits \cite{Bravyi2010, Breuckmann2021, yamasaki2024time, bravyi2024high}.

Superconducting qubits \cite{Nakamura1999, you_nature_2011, Xiang2013, Devoret_science_2013, krantz2019quantum, Blais2021}, one of leading platforms for fault-tolerant quantum computing, employ two primary types of connectivity.
The first is the all-to-all connected scheme, where each qubit can interact with any other \cite{majer2007, song2017, song2019}.
This scheme offers substantial benefits for entangling state preparations and quantum simulations \cite{song2017, song2019}.
However, its scalability is limited despite having achieved 20-qubit Sch\"{o}dinger cat states \cite{song2019}.
In contrast, the second method relies on nearest-neighbor interactions, as seen in the checker-board layout of solid-state quantum chips, compatible with the well-known surface code \cite{Fowler2012}.
This approach has enabled demonstrations of quantum advantages in specific problems and supported the scale up to hundreds of qubits \cite{arute_nature_2019, gong_science_2021, Wu_prl_2021} with the advancements in introducing variety but useful interactions \cite{Chen2014, yan_prapplied_2018, Xu_prl_2020, zhao_prl_2020, Sete2021, Sung_prx_2021}.
It has proven effective in quantum error correction \cite{google_quantum_ai_exponential_2021, krinner_realizing_2021} and quantum simulation \cite{Georgescu2014}.
Nevertheless, the restricted connectivity poses challenges to distributed computation and hardware-efficient error correction, prompting shifts towards enhancing this feature.

Some efforts have been made in establishing high connectivity recently \cite{Hazra2021, zhou2023realizing, McBroom2024}.
However, these proposals cannot avoid the challenges in scalability and hardware-efficient control.
An elegant scheme is to introduce long-range entangling gates for enhancing connectivity.
The first is the well-known teleported gate \cite{gottesman1999demonstrating}, that promises any distance two-qubit gates and scaling-up quantum architectures by utilizing modularity \cite{ang2022}.
Especially, the feasibility of achieving high-quality interconnections between quantum nodes using transmission lines \cite{zhong2019violating}, coaxial cables \cite{kurpiers2018deterministic, Campagne2018, zhong2021deterministic, niu2023low}, and waveguide \cite{Magnard2020,storz2023loophole,kannan2023demand} have been developed to achieve high-fidelity state transfers recently, while poor performance in quantum gate \cite{chou2018deterministic, niu2023low}.
This method to date has not been shown its advantages in experiments due to the limitations in gate speed, measurement performance, and coherence times.
From another view, many QECs do not need the ultra-long-range gate.
The second employs a novel strategy for facilitating coupling via a multimode cavity composed of a network of series-connected resonators \cite{McKay2015HighContrast, naik2017random}.
As the distance extends, the number of resonators increases linearly, indicating challenges in achieving connections between chips and among sample packages, thus constraining the scalability.
Hence, it is the necessity of a suitable long-range interaction for enhancing connectivity.

In this Letter, we introduce a novel scheme that employs two driving resonators and an interconnecting long-distance resonator,
stimulated by the multimode cavity \cite{McKay2015HighContrast}, to achieve strong $ZZ$ interaction between spatially separated transmons \cite{Koch2007} via the principle of the resonator-induced phase (RIP) gate \cite{Cross2015, Paik2016, Puri2016}.
The scheme is compatible with current fabrications,
and applies seamlessly into existing superconducting processors for implementing high coding-rate QECs that require enhanced connectivitys.
Through theoretical derivations and numerical simulations, we have analyzed the control parameters required.
Our findings indicate that high-fidelity CZ gates are achievable across a broad coupling distances.

\begin{figure}
    	\begin{minipage}[b]{0.5\textwidth}
    		\centering
    		\includegraphics[width=7.5cm]{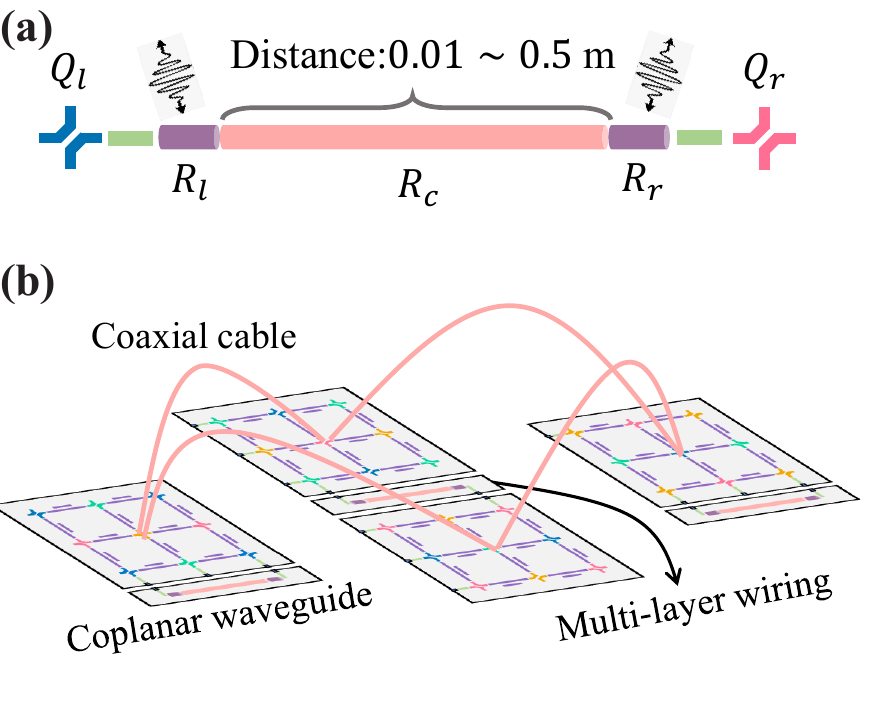}
    	\end{minipage}
		\caption{(Color online)
        (a) Schematic of the coupling architecture for long-range CZ gate.
            The left qubit $Q_l$ (blue) and the right qubit $Q_r$ (red) are interconnected via a multimode cavity comprised of a central resonator $R_c$ (pink) flanked by two auxiliary driving resonators $R_l$ and $R_r$ (purple) with resonant frequencies higher than those of the qubits.
            To facillitate $ZZ$ interactions, two microwaves are detunedly applied to the driving resonators.
            Additional green components indicate optional matching structures designed to refine coupling efficiency.
        (b) Diagram of a integrated superconducting processor with enhanced connectivity features.
            Within a single chip, extended long-range coupling is facilitated through a coplanar waveguide resonator, while intrachip connections are enabled based on multi-layer wiring in which the long-distance resonator can be deposited on a carrier chip.
            Interchip connections across different packages are linked using coaxial cables.
      \label{fig:FIG1}
      }
	\end{figure}

\textcolor{blue}{\textit{Theoretical model.}}---
    Our basic idea, as illustrated in Fig. \ref{fig:FIG1}(a), is using a long resonator, such as transmission line \cite{zhong2019violating} or coaxial cable \cite{kurpiers2018deterministic, Campagne2018, zhong2021deterministic, niu2023low}, which are universally applicable and easily controllable on quantum processors, to achieve long-range coupling between qubits.
    To facilitate precise driving of the long resonator and control their coupling with the qubits, we introduce two additional driving resonators with fundamental frequencies higher than that of the qubits to select the special high-order harmonic mode of the long resonator to form the multimode cavity following the work \cite{McKay2015HighContrast} where qubits interact through a network of strongly coupled resonators. 

    This is a simple but highly useful long-distance coupling architecture, as it can be directly implemented using now available fabrication processes and manipulation techniques.
    For example, as illustrated in Fig. \ref{fig:FIG1}(b), using flip-chip \cite{gold2021entanglement,Kosen2022} and multi-layer wiring \cite{yost2020} to achieve direct coupling between non-local qubits on the same chip or between chips, and non-local qubits between different chip packages can also be linked via coaxial cables \cite{niu2023low}.
    In our subsequent analysis, the findings suggest the advantages of our scheme for qubit separations ranging from approximately 0.01 to 0.5 m, where the distance between qubits can extend from $N \sim 5$ to $N \sim 250$, supposed a distance $d=2$ mm between nearest-neighbor qubits.
    Here, $N$ denotes the number of qubits spaced between two distant qubits.
    
    The long-range coupling potentially increases the connectivity of qubits.
    This enhancement is pivotal for reducing the overhead of coding rate in QECs, demonstrating practical implications for the implementation of quantum LDPC codes which require non-local couplings with additional long-range links \cite{bravyi2024high}.

    \begin{figure}
    	\begin{minipage}[b]{0.5\textwidth}
    		\centering
    		\includegraphics[width=7.0cm]{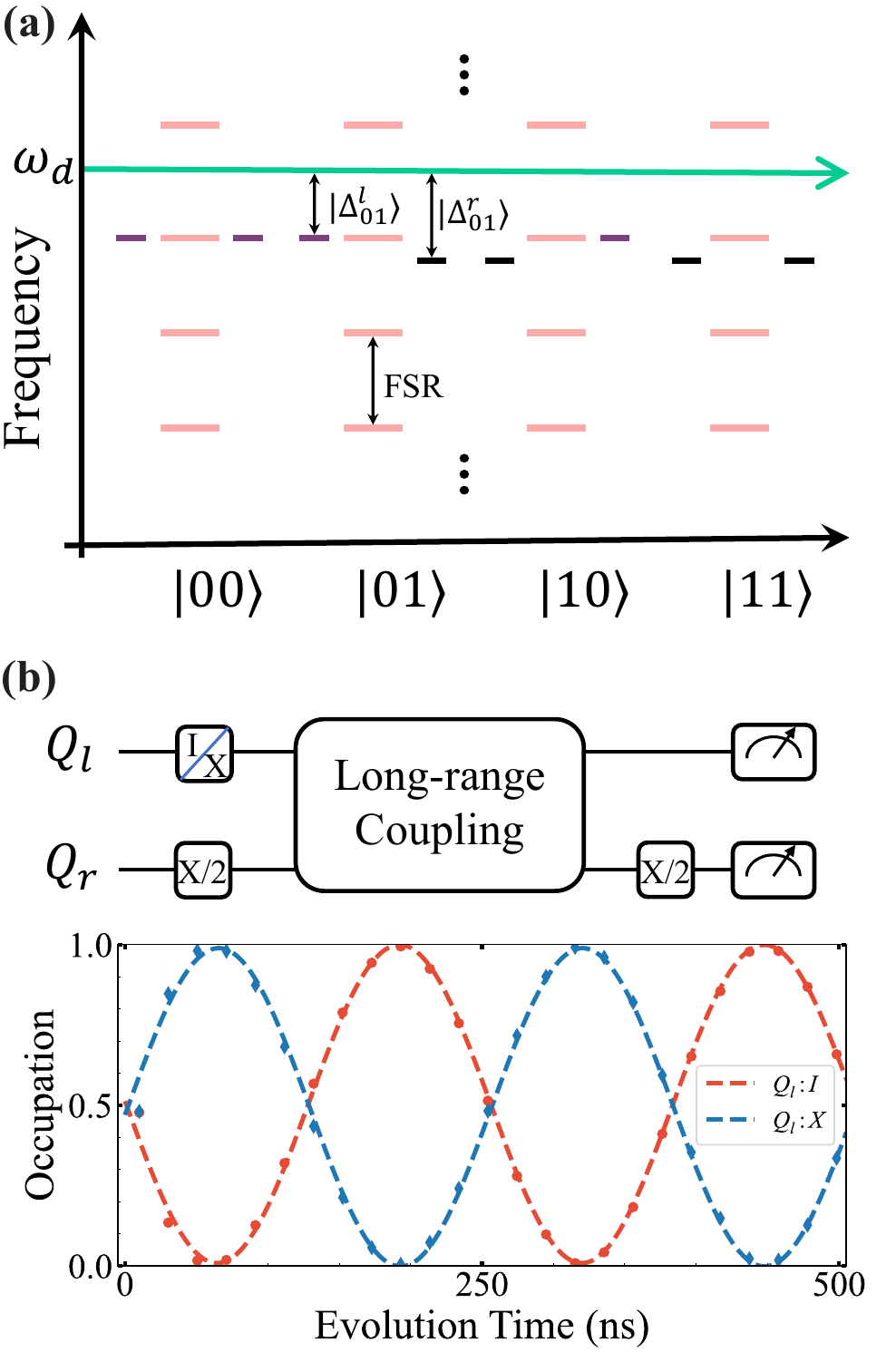}
    	\end{minipage}
    	\caption{
		(Color online) (a) Diagram illustrating long-range interactions mediated by the driving modes (in purple) and the high-order mode (in pink, featuring a small FSR) of the long-distance resonator.
        These modes depende on the states of the qubits.
        The frequencies of IQ drives, denoted by  $\omega_d$ (green line), modulate the qubit-state-dependent modes.
        Following this coupled mode-qubit evolution, the multimode system returns to the vacuum state, leaving the qubits decoupled from the cavity with an acquired entangling phase. 
        The detunings between the resonators and the drivings are represented by $\Delta_{01}^{l}$ and $\Delta_{01}^{r}$.
        (b) Illustrates a calibration method typical of conventional CZ gates, adapted for our scheme.
        This involves measuring the state occupation of the target qubit $Q_r$ under the control qubit $Q_l$ being in either the ground state (red) or excited state (blue).
        \label{fig:FIG2}
		}
	\end{figure}

    To demonstrate this idea,
    within the dispersive regime where $|\omega^q_{01}-\nu_n^p|\ll g^{(q,p)}$, and applying Schrieffer-Wolff transformation, we express the Hamiltonian of our reduced model as
    \begin{equation}\label{eq_H0}
    	\begin{aligned}	
        	H =& \sum_{q=r,l}\tilde{\omega}^{q}_{01}\sigma^{\dagger}_{q} \sigma_{q} + \sum_{p=r,l,c} \tilde{\nu}^{p} b^{\dagger}_{p}b_{p}\\
        	+&\sum_{q=p=r,l} \chi^{(q,p)}b^{\dagger}_{p}b_{p}\sigma^{\dagger}_{q} \sigma_{q}
        	+\sum_{p=r,l} g^{(p,c)} (b_{p}b^{\dagger}_{c}+b^{\dagger}_{p}b_{c})
    	\end{aligned}
    \end{equation}
    where $\sigma^{\dagger}_q$ and $\sigma_q$ are the raising and lowering operators for the $q$ qubit, respectively, while $b^{\dagger}_p$ and $b_p$ represent the creation and annihilation operators for the $p$ resonator.
    The term $\tilde{\omega}^{q}_{01}$ denotes the dressed frequency relative to the qubit frequency $\omega^{q}_{01}$,
    $\tilde{\nu}^p$ specifies the effective harmonic frequency of resonators,
    $\chi^{(q,p)}$ indicates the dispersive interaction between the $q$ transmon and $p$ resonator, 
    and $g^{(p,c)}$ quantifies the coupling strength between resonators.
    For more details, refer to Supplementary Material \cite{Supplementary}.
    As depicted in Fig. \ref{fig:FIG2}(a),
    the system evolution follows a qubit-state dependent trajectory in phase space initiated from the vacuum state of the multimode cavity by sequentially turning on and off in-phase and quadrature (IQ) drivings with frequencies $\omega_d$.

    In details, considering state space $\{|00\rangle, |01\rangle, |10\rangle, |11\rangle\}$  in a two-qubit system,
    we derive the time-dependent equation for the photon number $\alpha_{jk}^{p}$ corresponding to the qubit state $|jk\rangle$.
    This evolution is governed by the differential equation
    \begin{equation}\label{eq_photonnum}
    	\dot{\alpha}_{jk}^{p}=-(\tilde{\Delta}_{jk}^{p}\alpha_{jk}^{p}+\frac{i}{2}\tilde{\epsilon}^{p})+\sum_{p}g^{(p,c)}\alpha_{jk}^{c}
	\end{equation}
	where $\tilde{\Delta}_{jk}^{p}=-i(\Delta_{jk}^{p}+\chi_{jk}^{p})+\kappa^{p}/2$ is the effective detuning,
    $\Delta_{jk}^{p}$ corresponds to the detuning between the frequency of drivings and the cavity mode dressed by the qubit state $|jk\rangle$,
    and $\kappa^{p}$ and $\tilde{\epsilon}^{p}$ represent the energy decay rate and the driving amplitude, respectively.
	Then, the resonator-induced phase can be cast as
	\begin{equation}\label{eq_zz}
        \theta_{ZZ}=\mu_{11,00}-\mu_{10,00}-\mu_{01,00}
	\end{equation}
     derived from the time-dependent phase evolution 
	\begin{equation}
         \dot{\mu}_{jk,nm}(t)=-\sum_{p=r,l} (\chi_{jk}^p-\chi_{nm}^p)\alpha_{jk}^p\alpha_{nm}^{*(p)}
	\end{equation}
    with the dispersive shift $\chi_{jk}^p$ dependent on state $|jk\rangle$.

    It is noteworthy that enabling a single driving does not enable simultaneous dispersive coupling of qubits,
    it appears unlikely that photons in the cavity can induce an entanglement phase.
    However, in conditions where driving resonators are tuned to resonance with the specific mode of the long-distance resonator, the joint qubit-cavity evolution controlled by IQ drivings significantly depends on the qubit states.
    To elucidate this interaction, we reformulate the photon evolution described by Eq. \eqref{eq_photonnum} in vector written as $\dot{\vec{\alpha}}_{jk}=\hat{G}_{jk}\cdot \vec{\alpha}_{jk}+\vec{E}$.
    By applying a unitary $\hat{U}_{jk}$ to diagonalize the interaction matrix  $\hat{G}_{jk}$ into $\hat{D}_{jk}$, the transformed system dynamics are expressed as $\dot{\vec{\beta}}_{jk}=\hat{D}_{jk}\cdot \vec{\beta}_{jk}+\vec{E}^{'}_{jk}$.
    The solution is then provided by
	\begin{equation}
    	\beta_{jk}^p(t)=-\frac{i}{2}\int_0^t dt^{'} e^{-d_{jk}^p(t-t^{'})}E^{'}_{jk}
	\end{equation}
	where
    $E^{'}_{jk}$ denotes the diagonalizing modification of driving fields $\vec{E}$,
    $d_{jk}^p$ represents the diagonal elements of $\hat{D}_{jk}$.
    Therefore, adjustments to the driving amplitude, frequency, and length are strategically implemented to facilitate the realization of $ZZ$ interaction, thereby harnessing the long-range and high-fidelity CZ gate.
    
    Conventional CZ calibration techniques can be directly applied to our scheme.
    For instance, a standard calibration method we used, as illustrated in Fig. \ref{fig:FIG2}(b),
    involves the preparing of the target qubit $Q_r$ on a superposition state $|+\rangle = (|0\rangle + |1\rangle)/\sqrt{2}$, while the control qubit $Q_l$ is set in either the ground or excited states.
    To quantify the phase accumulated during these dynamics, 
    we project the state of the target qubit back onto $|+\rangle$ for measurement, allowing for obtaining the controlled phase by comparing the results across different states of $Q_l$.
    The corresponding evolution trajectories of $Q_l$ underscores a appreciable CZ gate operation time of $125$ ns.

    \begin{figure*}
        \begin{minipage}[b]{1.0\textwidth}
            \centering
            \includegraphics[width=15cm]{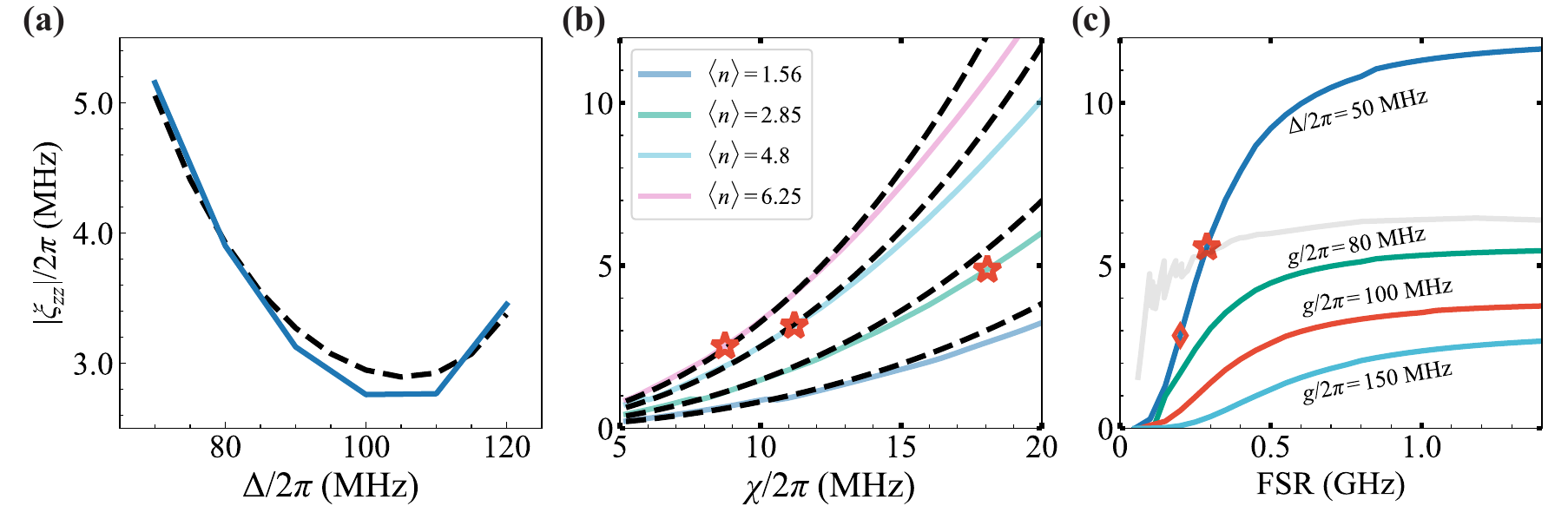}
        \end{minipage}
        \caption{
            (Color online)
            (a) The long-range $ZZ$ interaction is mediated by the driving detuning.
            The numerical simulations are in good agreement with theoretical predictions.
            Parameters are set at $\tilde{\epsilon}^l/2\pi=-\tilde{\epsilon}^r/2\pi=200$ MHz, $\chi^l/2\pi=20.12$ MHz, $\chi^r/2\pi=20.22$ MHz, and $g/2\pi=100$ MHz.
            (b) $ZZ$ interactions, from theoretical predictions (dashed lines) and simulations (solid lines), are influenced by the average photons and dispersive shifts.
            $ZZ$ interactions are acquired under average photons at 1.56, 2.85, 4.8, and 6.25, respectively.
            Star markers signify interactions where $\langle n \rangle $ touch the critical photons at corresponding dispersive shifts. 
            (c) The interaction varies with the FSR which is determined by the coupling distance. 
            The interaction strength varies inversely with FSR, converging to zero as FSR decreases. 
            Enhanced $ZZ$ interaction is achievable through optimization of coupling strength $g$ and detuning parameters.
            Specifically, at $g/2\pi = 80$ MHz and $\Delta/2\pi = 50$ MHz with an FSR of 200 MHz, a $ZZ$ interaction of 2.85 MHz is recorded (indicated by diamond marker).
            For comparative analysis, detuning is consistently set at $\Delta/2\pi = 80$ MHz for various coupling strengths $g/2\pi = 80$, $100$, and $150$ MHz, demonstrating varying interaction strengths.
            \label{fig:FIG3}
        }
    \end{figure*}

    \textcolor{blue}{\textit{Results.}}---
    First, we explore the dependence of $ZZ$ interaction on the detuning $\Delta$
    using the parameters $\tilde{\epsilon}^l/2\pi=-\tilde{\epsilon}^r/2\pi=200$ MHz, $\chi^l/2\pi=20.12$ MHz, $\chi^r/2\pi=20.22$ MHz, and $g/2\pi=100$ MHz,
    which support a $ZZ$ interaction, denoted as $\xi_{ZZ}$, exceeding 3 MHz, as shown in Fig. \ref{fig:FIG3}(a).
    Here, $g$ represents the coupling when $|g^{l,c}|/2\pi=|g^{r,c}|/2\pi$.
    Numerical simulations performed using QuTiP \cite{johansson2012qutip, Johansson_2013} validate our theoretical findings described by the formula Eq. \eqref{eq_zz}, demonstrating close alignment between the observed $ZZ$ interactions (blue dashed line) and our predictions (black solid line).
    The $ZZ$ interaction for the CZ gate in our scheme is approximately equivalent of that achieved with the prior RIP gate \cite{Paik2016}.
    Our approach enables the design of optimal parameters that achieve gate speeds of 100 ns with fidelities exceeding $0.99$.

    Next, the entanglement phase is significantly influenced by dispersive shifts $\chi_{jk}^{p}$.
    Under a specific IQ driving amplitude $\tilde{\epsilon}$, the real part of differentiating phase with time, viz. $ZZ$ interactions, can be expressed as
    \begin{equation}\label{eq_thetazz}
        \begin{aligned}
        \mathrm{Re}[\dot{\theta}_{zz}] \approx & \frac{\tilde{\epsilon}^2\bar{\chi}^2}{4\Delta}[\frac{9}{9\Delta^2-18g^2-6\Delta\bar{\chi}+\bar{\chi}^2}\\
        &-\frac{2}{2\Delta^2-3\Delta\bar{\chi}+\bar{\chi}^2}]
        \end{aligned}
    \end{equation}
    and the imaginary part is given by
    \begin{equation}\label{eq_coherence}
        \begin{aligned}
             \mathrm{Im}[\dot{\theta}_{zz}] \approx &\frac{\tilde{\epsilon}^2\bar{\chi}^2\kappa}{4\Delta}[\frac{4\Delta-3\bar{\chi}}{2(\Delta^2-3\Delta\bar{\chi}+\bar{\chi}^2)^2}\\
             &-\frac{27(3\Delta-\bar{\chi})}{(9\Delta^2-18g^2-6\Delta\bar{\chi}+\bar{\chi})^2}]
         \end{aligned}
    \end{equation}
    which outlines the decoherence rate due to the unstable state during photon injection into driving resonators.
    As shown in Fig. \ref{fig:FIG3}(b), a significant dispersion shift, with a large number of photons, ensures a robust $ZZ$ interaction.
    However, this introduces a trade-off concerning the critical photon, $n_{\text{crit}}=\left[(\delta+\eta)^2/4g^2-1\right]/3$, which sets an upper limit on the number of photons in the dispersive limit \cite{Blais2004}.
    Here, $\delta$ represents the detuning between the driving resonator and qubit, while the anharmonicity of the transmon is set as $\eta/2\pi = -0.28$ GHz in numerical caculation.
    We using star markers denote dispersion shifts corresponding to different critical photons.
    Notably, the average photons curve at 1.56 is not marked, as its corresponding dispersion shift surpasses 20 MHz.
    Our findings suggest that larger dispersion shifts, maintaining the same $ZZ$ interaction, facilitate meeting the dispersion condition more reliably.

    We further study the impact of FSR—equivalent to resonator length—on the $ZZ$ strength within the experimental framework.
    Fig. \ref{fig:FIG3}(c) demonstrates that a wide range of FSR values facilitates significant $ZZ$ interactions, characterized by $\tilde{\epsilon}^l/2\pi = \tilde{\epsilon}^r/2\pi = 200$ MHz, $\chi^l/2\pi=20.12$ MHz, $\chi^r/2\pi=20.22$ MHz.
    The interaction strength exhibits an inverse relationship with FSR, diminishing as FSR decreases.
    Optimization of the coupling strength $g$ and detuning $\Delta$ parameters can enhance $ZZ$ interaction.
    Notably, at $g/2\pi = 80$ MHz and $\Delta/2\pi = 50$ MHz with an FSR of 200 MHz, we observe a $ZZ$ interaction of 2.85 MHz, indicated by a diamond marker.
    The grey line describes the interaction as $\langle n \rangle $ approaches the critical photon for various FSR and detunings.
    A comparative analysis, maintaining $\Delta/2\pi = 80$ MHz while varying $g$ to 80, 100, and 150 MHz, reveals different interaction strengths.
    As FSR increases, the $ZZ$ strength stabilizes, facilitating a simplified computational model, thereby potentially reducing simulation overheads \cite{Supplementary}.

\begin{figure}
    	\begin{minipage}[b]{0.5\textwidth}
    		\centering
    		\includegraphics[width=7.0cm]{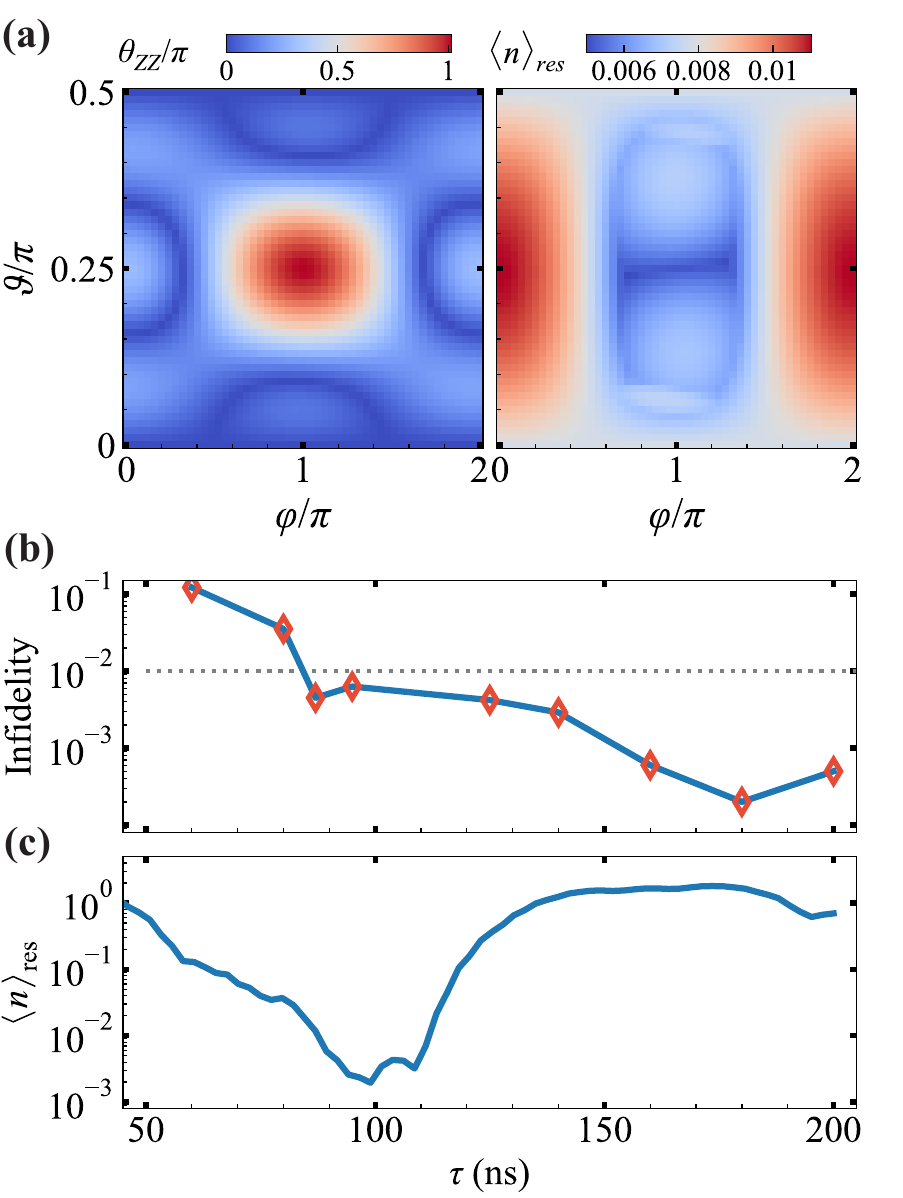}
    	\end{minipage}
    	\caption{
    		(Color online)
            (a) Explores the dependency of the entangling phase on IQ driving phase difference $\varphi$ and the coupling strength ratio $\vartheta = \tan^{-1}(|g^{l,c}|/|g^{r,c}|)$.
            Further elucidates the corresponding residual photons how influenced by these parameters.
            Demonstrates that optimal CZ gate performance is achieved under conditions of balanced coupling strengths $|g^{l,c}|=|g^{r,c}|$, with the phase difference $\varphi=\pi$ for the resonate mode number $m$ of the long-distance resonator that is even, and $\varphi=0$ when $m$ is odd.
            (b) Demonstrates the infidelity of the CZ gate with FSR=1.4 GHz across various evolution lengths, indicating a gate fidelity of 0.9925 for speeds exceeding 80 ns and an fidelity 0.999 within 160 ns, which is sufficient for implementing QECs.
            (c) Optimizes the driving pulses  with FSR=0.2 GHz across various evolution lengths, indicating a fast CZ gate can be realized with low residual photons $\langle n \rangle_{\text{res}} \sim 10^{-3}$ at $\tau=100$ ns.
    		\label{fig:FIG4}
    		}
    \end{figure}

    Moreover, it is crucial to ensure that photons are concentrated within the two driving resonators connected to the qubits.
    For clarity, we define a parameter $\vartheta = \tan^{-1}(|g^{l,c}|/|g^{r,c}|)$ to characterize the difference between resonators coupling, and introduce the difference in phase $\varphi$ between two driving microwaves.
    This framework benchmarks the optimal driving conditions for CZ gate where equal amplitudes and antiphase settings of the two drives, as illustrated in Fig. \ref{fig:FIG4}(a),
    which shows resilience to noise caused by experimental imperfections.
    Correspondingly, these conditions do not significantly increase the residual photons lead to exceede the threshold of critical photons.

    Therefore, as shown in Fig. \ref{fig:FIG4}(b), under optimized parameters, we assess the fidelity of the CZ gate and discover that theoretical fidelity remains below 99.9\% after a gate duration of 160 ns, and below 99\% after 80 ns.
    The primary error source is the residual photon, which becomes critical for gate times shorter than 80 ns, causing rapid fidelity degradation due to photon exceeding the critical photons.
    
    At last, we further validate our scheme apply to FSR=0.2 GHz achieving CZ gate with low residual photons and the speed nearly approaching 100 ns, details see Supplementary Materials \cite{Supplementary},
    adopted alongside varied driving pulses aimed at minimizing the residual photons.
    Following the fast adiabatic pulse proposed by Martinis and Geller in 2014 \cite{Martinis2014}, 
    we design an optimized waveform that successfully suppressed the residual photon to below 0.01 within approximately 100 ns.

    \textcolor{blue}{\textit{Discussions and conclusions.}}---
    We observed that the low-order harmonic mode of the long-distance resonator, proximal to the qubit and mediated by the driving resonator, facilitates effective $XX$ interaction.
    A detailed theoretical analysis, provided in the Supplementary Materials \cite{Supplementary}, identifies this as a significant error source when the FSR is minimal.
    To mitigate this, introducing direct couplings of qubits to the long-distance resonator \cite{Kandala2021} or employing tunable couplers \cite{yan_prapplied_2018,niu2023low} could suppress interactions between qubits and the resonator.
    Furthermore, to prevent qubit leakage into the multimode, the staggering interactions between the qubit and the mode of the long-distance resonator can be introduced.
    Setting $2g_{\text{eff}}/\text{FSR} \sim 0.1$ ensures that, under the condition $\langle n\rangle < n_{\text{crit}}$, the qubit state plays merely as an perturbation to the long resonator, effectively weaken leakages.
    
    Notably, our scheme consistently maintains negligible state occupation in the long-distance resonator \cite{Supplementary}. 
    This characteristic provides a distinct advantage in practical applications, particularly with current fabrication technologies.
    For example, it mitigates challenges such as low quality factors and thermalization issues stemming from the compromised performance of the resonator. 

    To conclude,
    we have successfully demonstrated the utility of the principle of RIP gates for executing long-range $ZZ$ interactions.
    Our results confirmed that our scheme yields beyond 99.9\% fidelity for a 160 ns CZ gate at FSR of 1.4 GHz, and further reduce residual photons reaching about $10^{-3}$ within 100 ns at FSR of 0.2 GHz.
    We believe our work provides a practical scheme for realizing high-fidelity and fast CZ gate across extended distances ranging from millimeters to sub-meters, thereby significantly enhancing quantum connectivity and enabling new QECs possibilities.

    ~\\ 
    \textit{Acknowledgements.}---
    We thank Junling Long for insightful discussion.
    This work was supported by
    NSFC (Grants No. U21A20436, and No. 12074179),
    Innovation Program for Quantum Science and Technology (Grant No. 2021ZD0301702),
    NSF of Jiangsu Province (Grants No. BE2021015-1, and BK20232002),
    Jiangsu Funding Program for Excellent Postdoctoral Talent (Grants No. 20220ZB16, and No. 2023ZB562),
    and 
    Natural Science Foundation of Shandong Province (Grant No. ZR2023LZH002).

    ~\\ 
    \textit{Note added.}---
    In preparing our manuscript, we noticed that two other studies reporting experiments on SWAP gates achieved through Raman transition \cite{mollenhauer2024} and CNOT gates achieved using cross resonance \cite{song2024} on superconducting devices interconnected by coaxial cables.

    \bibliography{reference}

    \onecolumngrid

    \newpage

    \setcounter{section}{1}
    \setcounter{equation}{0}
    \setcounter{page}{1}
    \renewcommand{\theequation}{S\arabic{equation}}
    \renewcommand{\thesection}{S\arabic{section}}
    \renewcommand{\thesubsection}{S\arabic{subsection}}
    \renewcommand{\thefigure}{S\arabic{figure}}
    \renewcommand{\thetable}{S\arabic{table}}

    \begin{center}
      {\large \textbf{Supplemental Material for 
      ``Long-Range $ZZ$ interaction via Resonator-Induced Phase in Superconducting Qubits"}}
    \end{center}


        ~\\
        ~\\
        ~\\
    
        In the Supplementary Material, we expand on the theoretical predictions and numerical simulations stated in main text, offering a deeper insight into the specifics of our proposed scheme.
        The first section elucidates the details of our theoretical model, followed by analytical results on the $ZZ$ interaction in the second section.
        The third section provides additional numerical simulation results, enhancing our understanding of the system dynamics under various conditions.
        The fourth section outlines proposed design parameters for superconducting qubit fabrication, illustrating the principles behind parameter design.
        The fifth section enriches our discussion with considerations of optimal control strategies.
        Our designs facilitate long-range, high-fidelity, and fast CZ gates, thereby improving qubit connectivity.
        This enhancement is crucial for experimental implementations of error-correcting codes like quantum LDPC codes that require extensive quantum connectivity and for achieving distributed quantum computing through long-range qubit coupling.
    
        ~\\
        ~\\
        ~\\
    \section{Theoretical model}
    
    \noindent	
    \textcolor{blue}{\textit{Motivation.}}---
    In recent years, using microwave resonators have enabled the preparation of high-fidelity entangled states over distances from meters to tens of meters, either through flying photons \cite{kannan2023demand} or direct coupling \cite{zhong2021deterministic, niu2023low}.
    These capabilities are crucial for scaling superconducting qubits and advancing distributed quantum computing.
    
    Our motivation is to achieve direct coupling between superconducting qubits across distances from millimeters to meters.
    There are two primary considerations: 
    First, achieving high coding rate in quantum error-correcting codes (QECs) demands increased quantum connectivity.
    For instance, one of quantum LDPC codes require each qubit to engage in high-fidelity two-qubit gates with at least six other qubits \cite{bravyi2024high}.
    Second, the long-range coupling we target, across distances from millimeters to meters, allows for direct realizing entanglement between qubits, facilitating fast and high-fidelity gates with existing sample fabrications.
    This is a plug-and-play part to current qubit coupling schemes and suits the scaling-up strategies of current distributed quantum computing. 
    Therefore, there is a motivation to develop long-range coupling to bridge the gap in sub-meter level coupling distances between qubits on superconducting chips.
    
    Our scheme begins by using a long-distance resonator (microwave photon carrier) as the medium for long-range coupling.
    However, as the resonator distance increases, its corresponding free spectral range (FSR) decreases, leading to the participation of other-order harmonic modes, which can significantly degrade two-qubit gate fidelity and operation speed.
    Thus, accurately selecting specific high-order harmonic modes is critical.
    
\begin{figure*}[htbp]
        \begin{minipage}[b]{1.0\textwidth}
            \centering
            \includegraphics[width=15.0cm]{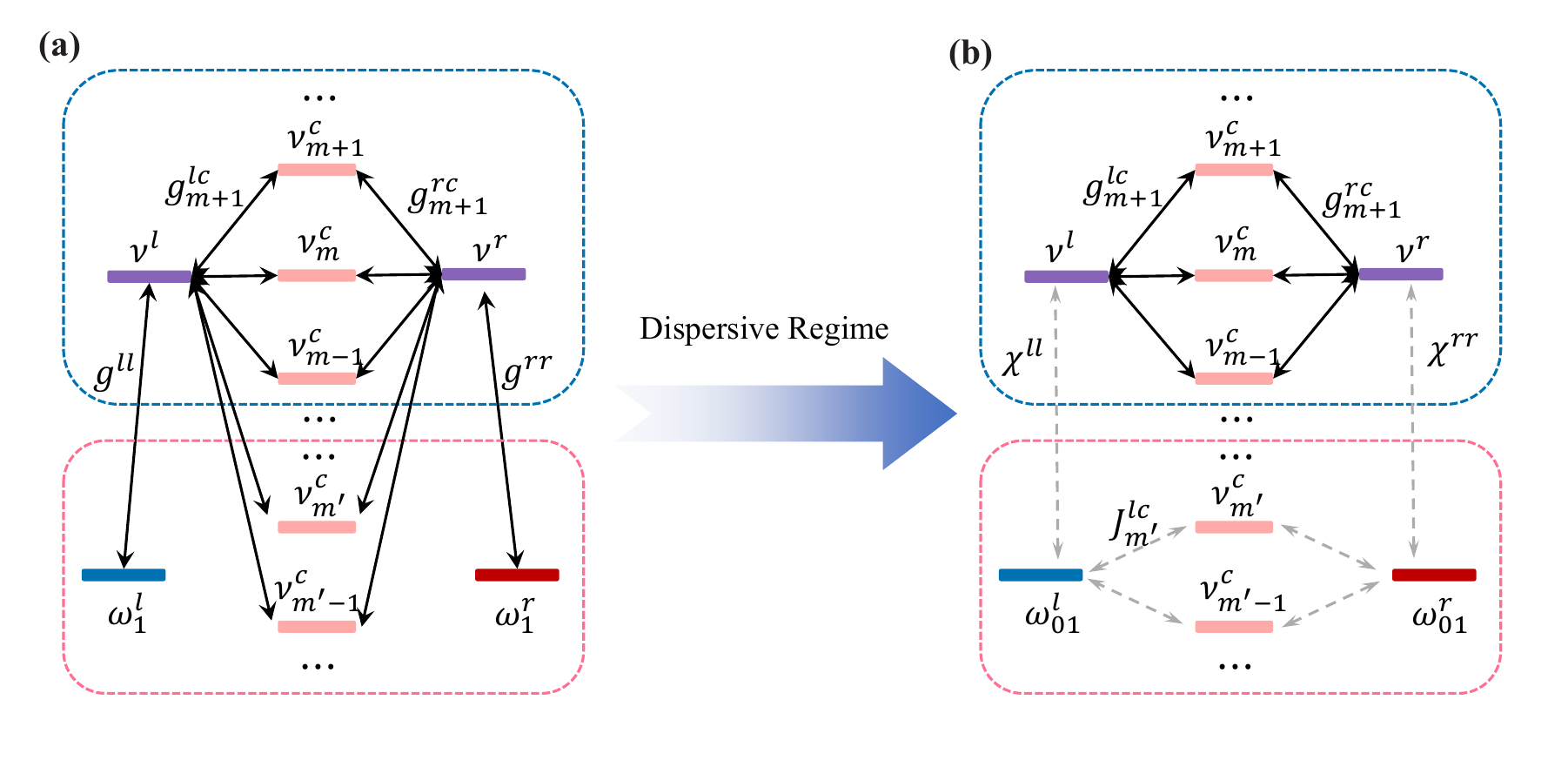}
        \end{minipage}
        \caption{(Color online)
        Energy diagram of the theoretical framework.
        (a) The model incorporates two qubits characterized by a transition frequency $\omega_{01}^{q}$ and weak anharmonicity $\eta^{q}$, alongside two phase-modulated driving resonators with a resonance frequency $\nu^{p}$ surpassing $\omega_{01}^{q}$, where $q=p=r,l$.
            A long-distance resonator, described by a free spectral range (FSR) with harmonically distributed modes $\nu_m^{c}$, facilitates long-range coupling between spatially separated qubits.
            Interaction strengths $g^{p,q}$ and $g_m^{p,c}$ represent the $XX$ interactions within this setup.
            The model delineates two principal regions:
            the high energy region (HER) and the low energy region (LER), depicted respectively in blue and pink.
        (b) In the dispersive regime, where $\Delta^{p,q} \gg g^{p,q}$, dispersive shifts $\chi^{p,q}$ emerge, with detuning $\Delta^{p,q}$ defined as $\omega_{01}^{q}-\nu^{p}$.
            This regime facilitates effective $XX$ couplings $J_m^{p,q}$ between qubits and LER modes of the long-distance resonator, mediated by the driving resonator.
            Special attention is required to suppress these interactions to prevent leakage of the qubit states into the resonators.
            In HER, if $\textrm{FSR} \gg g^{p,c}$, the tri-resonator system forms a dark state with zero eigen-energy, akin to the reduced model depicted in Fig. \ref{fig:smFIG2}.
            In this regime, the RIP gate mechanism demonstrates that photon dynamics influenced by qubit states are confined to interactions between driving resonators, indicating robustness against defects in the long-distance resonator.
            It is noteworthy that, when $\textrm{FSR} \gg g^{p,c}$ is not met, the model can also be equivalent to a harmonic mode facilitating $ZZ$ interactions, though the strength of these interactions diminishes as FSR decreases.
            This necessitates careful consideration of optimal parameter design for varying FSR levels, as discussed in Sec. 4, we list some parameters for different FSR.
      \label{fig:smFIG1}
      }
    \end{figure*}

    To address this, we introduce additional resonators as `filters', following the work to achieve multimode coupling between qubits through a network of strongle coupled resonators \cite{McKay2015HighContrast},
    to select the desired order modes for inducing coupling, as depicted in Fig. \ref{fig:smFIG1}.
    But only this, the coupling is inherently weak.
    Stimulated by the resonator-induced phase (RIP) scheme \cite{Cross2015, Paik2016, Puri2016},
    we enhance coupling strength by applying two phase-modulated microwave drives to  `filters' as driving resonators, thereby achieving a strong $ZZ$ interaction.
    This method circumvents the practical challenges of applying drives into long-distance resonator, and allows for the adjustment of driving parameters, such as phases and amplitudes, to achieve coherent cancellation and hence acquiring a cleaner two-qubit coupling.
    
    ~\\
    \noindent	
    \textcolor{blue}{\textit{The Hamiltonian of the model.}}---
    Without loss of generality, we introduce the basic model as depicted on the left of Fig. \ref{fig:smFIG1}, comprising five components.
    The corresponding Hamiltonian is given by
    \begin{equation}\label{smeq:H_all}
        \begin{aligned}
            H = & \sum_{q=r,l} \omega^{(q)}_{01}\sigma_q^{\dagger}\sigma_q+\frac{\eta^{(q)}}{2}\sigma_q^{\dagger}\sigma_q^{\dagger}\sigma_q\sigma_q + \sum_{p=r,l} \nu^{(p)} b^{\dagger}_{p}b_{p} +  \sum_m \nu_m^{c} b^{\dagger}_{c,m}b_{c,m}\\
                & +\sum_{q=p=r,l} g^{(q,p)}(\sigma_q b^{\dagger}_{p}+\sigma_q^{\dagger} b_{p})\\
                & +\sum_m g_m^{lc}(b_{l} b^{\dagger}_{c,m}+b^{\dagger}_{l} b_{c,m}) +\sum_m (-1)^{m} g_m^{rc}(b_{r} b^{\dagger}_{c,m}+b^{\dagger}_{r} b_{c,m})\\
        \end{aligned}
    \end{equation}
    where $\sigma^{\dagger}$ ($b^{\dagger}$) and $\sigma$ ($b$) denote the raising and lowering operators for qubits (resonators),
    $\omega^{(q)}$ and $\eta^{(q)}$ represent the frequency and weak anharmonicity of the $q$-th qubit, respectively, while $\nu^{(p)}$ corresponds to the frequency of the $p$-th resonator,
    The interaction strengths between the $q$-th qubit and the $p$-th resonator, and between the driving resonator and the $m$-th mode of the long-distance resonator, are denoted by $g^{(q,p)}$ and $g_m^{lc}$, respectively,
    Note that the sign of $g_m^{rc}$ alternates with mode number $m$, reflecting the parity changes in the standing wave mode \cite{Pellizzari1997, Vogell2017}.

    To analyze this model, we categorize the modes of the long-distance resonator into two types:
    those close to the frequencies of both side resonators (denoted high-energy region) and those close to the frequencies of qubits (denoted low-energy region).
    This categorization necessitates the fulfillment of the following assumptions
    \begin{equation}\begin{aligned} \label{smeq:EqS20}
        \left|\omega^{(q)} - \nu^{(p)}\right| \gg g^{(q,p)} \gg g^{(p,c)}
    \end{aligned}\end{equation}
    
    Furthermore, under the SWT and the first-order approximation, we can obtain the effective Hamiltonian as follows
    \begin{equation}\begin{aligned} \label{smeq:H_all_disp}
        H^{'} =& \sum_{q=r,l}\frac{\tilde{\omega}^{(q)}}{2} \sigma^{z}_q + \sum_{p=r,l} \nu^{(p)} b^{\dagger}_{p}b_{p} +\sum_{m}\mu^{(c)}_m b_{c,m}^{\dagger}b_{c,m} \\
                 & +\sum_{q=p=r,l} \chi^{(q,p)} b_{p}^{\dagger}b_{p}\sigma_q^{\dagger}\sigma_q  + \sum_{q=r,l} \sum_{m} J_{m}^{(q,c)}\left(b_{c,m}^{\dagger}\sigma_q + b_{c,m}\sigma_q^{\dagger}\right)\\
                 & +\sum_m g_m^{lc}(b_{l} b^{\dagger}_{c,m}+b^{\dagger}_{l} b_{c,m}) +\sum_m (-1)^{m} g_m^{rc}(b_{r} b^{\dagger}_{c,m}+b^{\dagger}_{r} b_{c,m})\\
    \end{aligned}\end{equation}
    with the frequency of the dressed state defined as
    \begin{equation}\label{smeq:EqS3}
        \tilde{\omega}^{(q)} = \omega^{(q)} + \frac{(g^{(q,p)})^2}{\omega^{(q)}-\nu^{(p)}}
    \end{equation}
    and the dispersive shift $\chi^{(q,p)} = \chi^{(q,p)}_1-\chi^{(q,p)}_0$ with
    \begin{equation}\label{smeq:EqS4}
        \chi^{(q,p)}_n = \frac{(g^{(q,p)})^2(\eta^{(q)}-\omega^{(q)}+\nu^{(p)})}{(\omega^{(q)}+n\eta^{(q)}-\nu^{(p)})(\omega^{(q)}+(n-1)\eta^{(q)}-\nu^{(p)})}
    \end{equation}
    where $n$ corresponds to the qubit state $|n\rangle$. 
    
    And the effective coupling between qubits can be written as
    \begin{eqnarray} \label{smeq:EqS22}
        J_{m}^{(q,c)} = \frac{g^{(q,p)}g^{(p,c)}}{2}\left[\frac{1}{\tilde{\omega}^{(q)}-\nu^{(p)}}+\frac{1}{\nu_m^{(c)}-\nu^{(p)}}-\frac{1}{\tilde{\omega}^{(q)}+\nu^{(p)}}-\frac{1}{\nu_m^{(c)}+\nu^{(p)}}\right]
    \end{eqnarray}

    In the design of practical samples, it is crucial to suppress the leakage-inducing effective $XX$ coupling while enhancing the dispersive shift to facilitate the $ZZ$ interaction.
    Strategies to mitigate $XX$ interactions include the introduction of direct couplings between qubits and the long-distance resonator \cite{Kandala2021} or the use of tunable couplers \cite{yan_prapplied_2018, niu2023low}, which can effectively minimize interactions between qubits and the resonator.
    Additionally, to prevent qubit leakage into the multimode, staggered interactions between the qubit and the modes of the long-distance resonator may be employed.
    Setting the ratio $2J_{m}^{(p,q)}/\text{FSR} \sim 0.1$ ensures that, under the dispersive regime, the qubit state merely acts as a perturbation to the long-distance resonator, thereby effectively reducing leakages.

    \begin{figure*}[htbp]
        \begin{minipage}[b]{1.0\textwidth}
            \centering
            \includegraphics[width=15.0cm]{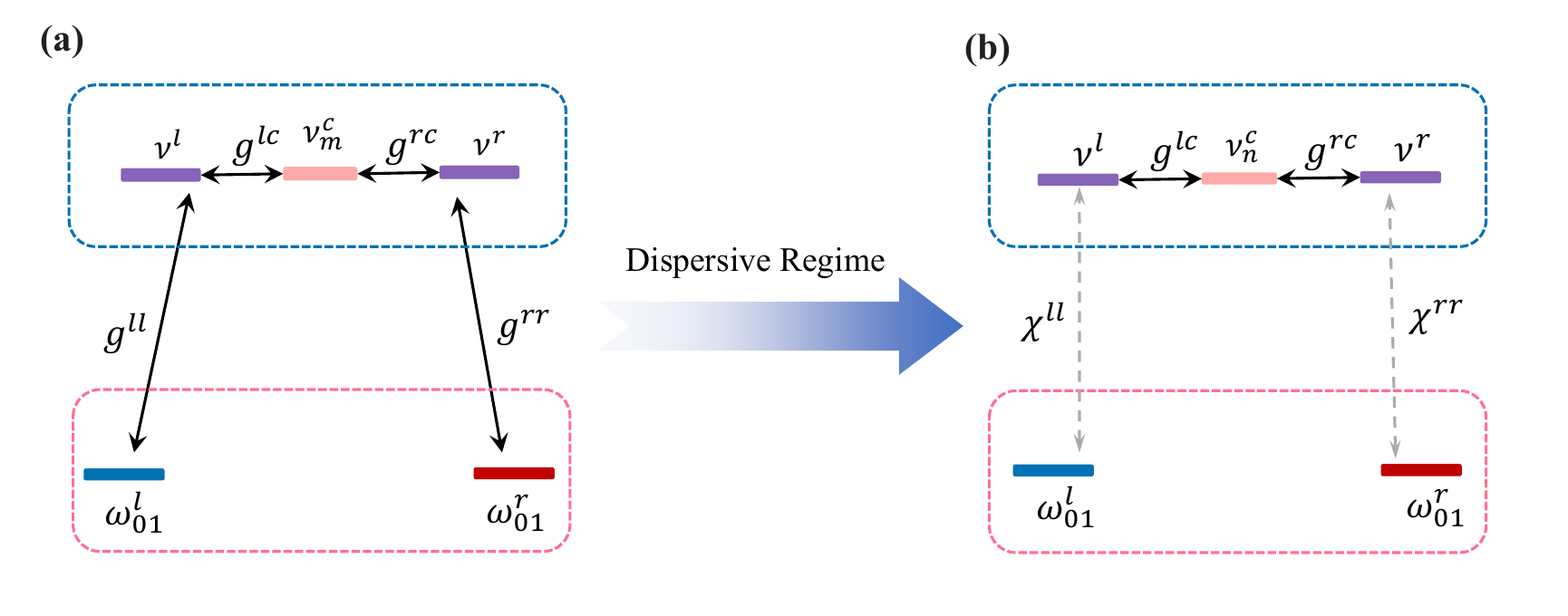}
        \end{minipage}
        \caption{(Color online)
        Energy diagram of the reduced model.
        (a) This model insightfully clarifies the foundational mechanisms of the theoretical model.
            It features two qubits with distinct frequencies $\omega^{r,l}$ that are substantially detuned from the driving resonators, denoted by $\nu^{l,r}$.
            These resonators align with a central harmonic mode $\nu_m^{c}$ within a long-range resonator.
        (b) In the dispersive regime, significant detuning between the qubits and driving resonators induces shifts in the dispersive coupling.
            A multimode structure arises from three serially coupled resonator modes, via coupling strengths $g^{lc}$ and $g^{rc}$, enabling strong coherent interactions among the driving resonators.
            These interactions correspond to an eigen-energy of zero.
            That means when photons are injected by drivings, indicating no photon occupancy in the central mode during the dynamics.
            Consequently, a change in the state of one qubit alters the photon dynamics in its corresponding driving resonator, affecting the photon dynamics in the opposite resonator, and ultimately leading to an accumulated controlled phase in the state of the other qubit.
      \label{fig:smFIG2}
      }
    \end{figure*}

    ~\\
    \noindent	
    \textcolor{blue}{\textit{Single mode selected.}}---
    Firstly, supposed $\textrm{FSR} \gg  g_m^{p,c}$, the model can reduce to the only one mode, denoted as $\nu^{c}$ for simplicity, of the long-distance resonator selected, as shown in Fig. \ref{fig:smFIG2} (a).
    The Hamiltonian Eq. \eqref{smeq:H_all_disp} reduce to 
    \begin{equation}\begin{aligned} \label{smeq:H_disp}
        H =& \sum_{q=r,l}\frac{\tilde{\omega}^{(q)}}{2} \sigma^{z}_q + \sum_{p=r,l,c} \nu^{(p)} b^{\dagger}_{p}b_{p}\\
            & +\sum_{q=p=r,l} \chi^{(q,p)} b_{p}^{\dagger}b_{p}\sigma_q^{\dagger}\sigma_q +\sum_{p=r,l} g^{(p,c)}(b_{p} b^{\dagger}_{c}+b^{\dagger}_{p} b_{c})
    \end{aligned}\end{equation}
    Here, we assume that the coupling between resonators has the same polarity and that the selected mode for long-distance interaction is even.
    
    It should be noted that in Eq. \eqref{smeq:H_disp}, terms corresponding to static $XX$ and $ZZ$ couplings are omitted, as these represent higher-order small quantities.
    For simplicity, we set $g_1 = g^{rr} = g^{ll}$, $g_2 = g^{rc} = g^{lc}$, and $\nu = \nu^{r} = \nu_m^{c} = \nu^{l}$ and $\omega = \omega^{r}_{01} = \omega^{l}_{01}$.
    Specifically, these terms can be expressed as follows
    \begin{eqnarray*}
        J_{xx}^{s} &=&  \frac{g_1^2 g_2^2}{(\omega-\nu)^3}\\
        \xi_{zz}^{s} &=& \frac{12(J_{xx}^{s})^2}{\omega-\nu}
    \end{eqnarray*}
    
    Based on the mechanism of the RIP scheme \cite{Cross2015},
    the state of the left qubit influences the resonator response to the driving field, while the frequency of the right qubit is modulated by photons within the corresponding harmonic mode.
    Consequently, the frequency of the right qubit depends on the state of the left qubit, and vice versa.
    In our model, we found this mechanism is preserved by applying two fields into the resonators connected to each qubit under in-phase and quadrature (IQ) drivings, that is $|\phi^{r}-\phi^{l}|=\pi$.
    These fields are identical, differing only in phase, allowing the three resonators to resonate synchronously, as stated in main text.
    Note that if the number of the mode of the long-distance resonator is odd, the driving phases satisfy $|\phi^{r}-\phi^{l}|=0$.

    ~\\
    \noindent	
    \textcolor{blue}{\textit{Two phase-modulated drivings.}}---
    Without loss of generality, we introduce two phase-modulated drivings with frequency $\omega^{p}_d$  and apply a frame transformation 
    $$
    R(t) = \exp{\left[-it\left(\sum_q\frac{\tilde{\omega}^{(q)}}{2}\sigma^{z}_q + \sum_p \omega_{d}^{p} b^{\dagger}_{p}b_{p}\right)\right]}
    $$
    into the reduced model, we then derive the model with drivings in the rotating frame as
    \begin{equation}\label{smeq:H_disp_driv}
        \begin{aligned}
            H^R = & \sum_{p=r,l}\Delta^{p} b^{\dagger}_{p}b_{p} + \sum_{q=p=r,l}\chi^{(q,p)}b^{\dagger}_{p}b_{p}\sigma^{\dagger}_{q} \sigma_{q}
                  + \sum_{p=r,l} g^{(p,c)}(b_{p} b^{\dagger}_{c}+b^{\dagger}_{p}b_{c})\\
                  & +\sum_p\frac{1}{2} \epsilon^p(t) \left[e^{i\phi^p}b^{\dagger}_{p} + h.c.\right]
        \end{aligned}
    \end{equation}
    where $\epsilon^p (t)$ represent the envelope of the driving in $p$-th resonator, and $\phi^p$ denotes the driving phase.
    The term $\Delta^{p} = \nu - \omega_d^{p}$ indicates the detuning between the resonators.

    We only consider the dynamics of the qubits within the computational subspace, we can write the density operator for the coupled system as
    $$
    \rho=\sum_{jk,lm}C_{jk,lm}\otimes|jk\rangle\langle lm|
    $$
    where $j, k, l$ and $m$ label states of qubits and $C_{jk,lm}$ are blocks of the density operator for three resonators.
    And the master equation can be written as
    \begin{equation}\label{smeq:rho_masterEq}
        \dot{\rho}=-i[H^R,\rho]+\sum_p\kappa^{(p)} D[b_p]\rho
    \end{equation}
    where $\kappa^{(p)}$ is the decay rate of the $p$-th resonator with decoherence operator $D[b_p]$.

    \section{Theoretical solutions}
    
    \noindent	
    \textcolor{blue}{\textit{The dynamics of the photon number.}}---
    Introducing a generalized P representation, the evolution equation of the photon in resonators can be obtained
    \begin{equation}\label{smeq:photon_dynamics}
        \left\{
            \begin{aligned}
            \dot{\mu}_{jk, lm}(t)=-\sum_{p}\left(\tilde{\chi}_{jk}^{(p)}-\tilde{\chi}_{l m}^{(p)}\right) \alpha_{jk}^{(p)} \alpha_{lm}^{(p)} \\ 
            \dot{\alpha}_{j k}^{(p)}=-\left(\tilde{\Delta}_{j k}^{(p)} \alpha_{j k}^{(p)}+\frac{i}{2} \tilde{\epsilon}^{p}\right)+ g^{(p,c)} \alpha_{jk}^{c}
        \end{aligned}
        \right.
    \end{equation}
    where $\tilde{\Delta}_{j k}^{(p)} = i\left(\Delta_{d}^{(p)}+\tilde{\chi}_{j k}^{(p)}\right)+\kappa^{(p)}/2$,
    and $\tilde{\chi}_{jk}^{(p)}=\chi_{j}^{(p)}+\chi_{k}^{(p)}$ is the sum of dispersive shifts caused by the left qubit and right qubit. 
    
    Here, we prove the formula Eq. \eqref{smeq:photon_dynamics} with a simple mathematical operation.
    Firstly, after substituting density operator into the master equation Eq. \eqref{smeq:rho_masterEq}, we can obtain the diagonal term
    \begin{equation}\begin{aligned} \label{smeq:EqS7}
        \dot{C}_{j k, j k} = & -i\sum_{p}\left(\Delta^{(p)} + \chi_{j k}^{(p)}\right)\left[b_{p}^{\dagger} b_{p}, C_{j k, j k}\right] \\
                            & -i\sum_{p}\sum_{p'>p}g^{(p',p)}[b_{p}b^{\dagger}_{p'}+b^{\dagger}_{p}b_{p'}, C_{j k, j k}]\\
                            &-\frac{i}{2} \sum_{p}\left[\tilde{\epsilon}_{p}^{*}(t)b_{p}+ \tilde{\epsilon}_{p}(t)b_{p}^{\dagger}, C_{j k, j k}\right] \\
                             &+ \sum_{p} \kappa_{p} D[b_{p}] C_{j k, j k}
    \end{aligned}\end{equation}
    and nondiagonal term
    \begin{equation}\begin{aligned} \label{smeq:EqS8}
        \dot{C}_{j k, l m} = & -i\sum_{p} \Delta^{(p)}\left[b_{p}^{\dagger} b_{p}, C_{j k, l m}\right] - i\sum_{p} \chi_{j k}^{(p)} b_{p}^{\dagger} b_{p} C_{j k, l m} + i\sum_{p} \chi_{l m}^{(p)} b_{p}^{\dagger} b_{p} C_{j k, l m} \\
                             & -\frac{i}{2} \sum_{p}\left[\tilde{\epsilon}_{p}^{*}(t)b_{p}+ \tilde{\epsilon}_{p}(t)b_{p}^{\dagger}, C_{j k, l m}\right] \\
                             & - i\sum_{p}\sum_{p'}g^{(p',p)}[b_{p}b^{\dagger}_{p'}+b^{\dagger}_{p}b_{p'}, C_{j k, l m}]\\
                             & + \sum_{p} \kappa_{p} D[b_{p}] C_{j k, l m}
    \end{aligned}\end{equation}
    
    Secondly, we make use of a generalized P representation
    $C_{jk,lm} = \int d\alpha^2 \int d\beta^2 \varLambda(\alpha,\beta)P_{jk,lm}(\alpha,\beta)$
    where
    $\varLambda(\alpha,\beta) =|\alpha\rangle\langle \beta^{*}|/\langle \beta^{*}|\alpha\rangle$,
    which gives the operator correspondences
    \begin{equation}
        \begin{aligned} \label{smeq:EqS9}
        b C_{j k, l m} & \leftrightarrow \alpha P_{j k, l m}(\alpha, \beta), \\ 
        b^{\dagger} C_{j k, l m} & \leftrightarrow\left(\beta-\partial_{\alpha}\right) P_{j k, l m}(\alpha, \beta), \\ 
        C_{j k, l m} b^{\dagger} & \leftrightarrow \beta P_{j k, l m}(\alpha, \beta), \\
        C_{j k, l m} b & \leftrightarrow\left(\alpha-\partial_{\beta}\right) P_{j k, l m}(\alpha, \beta) .
        \end{aligned}
    \end{equation}
    Then, we obtain commutation relations as follows
    \begin{align*}
        \Delta^{(p)}\left[b_{p}^{\dagger} b_{p}, C_{j k, l m}\right] 
        &= \Delta^{(p)} \left[\left(\beta_{p}-\partial_{\alpha_p}\right)\left(\alpha_{p} P_{j k, l m}\right)-\left(\alpha_{p}-\partial_{\beta p}\right)\left(\beta_{p} P_{j k, l m}\right)\right] \nonumber \\
        &= \Delta^{(p)} \left(-\alpha_{p}\partial_{\alpha_{p}} P_{j k, l m}+\beta_{p}\partial_{\beta_{p}} P_{j k, l m}\right) \nonumber \\
        \\ g^{(p', p)}\left[b_{p}^{\dagger} b_{p'}+b_{p} b_{p'}^{\dagger}, C_{j k, l m}\right] 
        &= g^{(p', p)}\left[\left(\beta_{p'}-\partial_{\alpha_{p'}}\right)\left(\alpha_{p} P_{j k, l m}\right)+\alpha_{p'}\left(\beta_{p}-\partial_{\alpha_{p}}\right) P_{j k, l m} \right. \nonumber \\
        & \quad \left. -\left(\alpha_{p}-\partial_{\beta_{p}}\right)\left(\beta_{p'} P_{j k, l m}\right)+\beta_{p}\left(\alpha_{p'}-\partial_{\beta_{p'}}\right) P_{j k, l m}\right] \nonumber \\ 
        &= g^{(p', p)}\left[\left(-\alpha_{p} \partial_{\alpha_{p'}}-\alpha_{p'} \partial_{\alpha_{p}}\right) P_{j k, l m}+\left(\beta_{p'} \partial_{\beta_{p}}+\beta_{p} \partial_{\beta_{p'}}\right) P_{j k, l m}\right] \nonumber \\
        \\ \chi_{j k}^{(p)} b_{p}^{\dagger} b_{p} C_{j k, l m}-\chi_{l m}^{(p)} C_{j k, l m} b_{p}^{\dagger} b_{p} 
        &= \left[\chi_{j k}^{(p)}-\chi_{l m}^{(p)}\right] \alpha_{p} \beta_{p} P_{j k, l m}-\chi_{j k}^{(p)} \partial_{\alpha_{p}}\left(\alpha_{p} P_{j k, l m}\right) +\chi_{l m}^{(p)} \partial_{\beta_{p}}\left(\beta_{p} P_{j k, l m}\right) \nonumber \\
        \\ \left[\tilde{\epsilon}_{p} b^{\dagger}+\tilde{\epsilon}_{p}^{*} b, C_{j k, l m}\right] 
        &= \tilde{\epsilon}_{p}\left(\beta_{p}-\partial_{\alpha_{p}}\right) P_{j k, l m}+\tilde{\epsilon}_{p}^{*} \alpha_{p} P_{j k, l m}- \tilde{\epsilon}_{p} \beta_{p} P_{j k, l m}-\tilde{\epsilon}_{p}^{*}\left(\alpha_{p}-\partial_{\beta_{p}}\right) P_{j k, l m} \nonumber \\ 
        &= -\tilde{\epsilon}_{p} \partial_{\alpha_{p}} P_{j k, l m}+\tilde{\epsilon}_{p}^{*} \partial_{\beta_{p}} P_{j k, l m} \nonumber \\
        \\ \kappa^{(p)} D\left[b_{p}\right] C_{j k, l m} 
        &= \frac{\kappa^{(p)}}{2}\left(2 b_{p} C_{j k,l m} b_{p}^{\dagger}-b_{p} b_{p}^{\dagger} C_{j k, l m}-C_{j k, l m} b_{p} b_{p}^{\dagger}\right) \nonumber \\
        &= \frac{\kappa^{(p)}}{2}\left[2 \alpha_{p} \beta_{p} P_{j k, l m}-\alpha_{p}\left(\beta_{p}-\partial_{\alpha_{p}}\right) P_{jk, l m}-\beta_{p}\left(\alpha_{p}-\partial_{\beta_{p}}\right) P_{j k, l m}\right] \nonumber \\
        &= \frac{\kappa^{(p)}}{2}\left(\alpha_{p} \partial_{\alpha_{p}}+\beta_{p} \partial_{\beta_{p}}\right) P_{j k, l m}
    \end{align*}

    Organize the above formulas to obtain
    \begin{equation}\label{smeq:Pjkjk_dt}
        \begin{aligned}
    \dot{P}_{j k,j k} = &
    \sum_{p} \left[\tilde{\Delta}_{jk}^{(p)} \alpha_{p}+\frac{i}{2} \tilde{\epsilon}_{p}(t)\right] \partial_{\alpha_p} P_{j k, j k}+\sum_{p} \sum_{p'>p} \tilde{g}^{(p,p')}\left[\left(\alpha_{p} \partial_{\alpha_{p'}}+\alpha_{p'} \partial_{\alpha_{p}}\right) P_{jk, jk}\right] \\ +& \sum_{p}\left[\tilde{\Delta}_{jk}^{(p)*} \beta_{p}-\frac{i}{2} \tilde{\epsilon}_{p}^{*}(t)\right] \partial_{\alpha_p} P_{jk,jk}+ \sum_{p} \sum_{p'>p} \widetilde{g}^{(p,p')*}\left[\left(\beta_{p} \partial_{\beta_{p'}}+\beta_{p'} \partial_{\beta_{p}}\right) P_{j k, j k}\right] \\
    \end{aligned}\end{equation}
    
    \begin{equation}\label{smeq:Pjklm_dt}
        \begin{aligned}
    \dot{P}_{j k,l m} = &
    \sum_{p} \left[\tilde{\Delta}_{jk}^{(p)} \alpha_{p}+\frac{i}{2} \tilde{\epsilon}_{p}(t)\right] \partial_{\alpha_p} P_{j k, l m}+\sum_{p} \sum_{p'>p} \tilde{g}^{(p,p')}\left[\left(\alpha_{p} \partial_{\alpha_{p'}}+\alpha_{p'} \partial_{\alpha_{p}}\right) P_{jk, lm}\right] \\ +& \sum_{p}\left[\tilde{\Delta}_{lm}^{(p)*} \beta_{p}-\frac{i}{2} \tilde{\epsilon}_{p}^{*}(t)\right] \partial_{\alpha_p} P_{jk,lm}+ \sum_{p} \sum_{p'>p} \widetilde{g}^{(p,p')*}\left[\left(\beta_{p} \partial_{\beta_{p'}}+\beta_{p'} \partial_{\beta_{p}}\right) P_{j k, lm}\right] \\-&i(\chi_{jk}^{(p)}-\chi_{lm}^{(p)}\alpha_{p}\beta_{p}P_{jk,lm})
    \end{aligned}\end{equation}
    
    where $\tilde{g}^{p,p'}=i g^{p,p'}$.
    Additionally, in the dispersive regime, we have  
    \begin{equation}\label{smeq:assum_P}
        \begin{aligned}
            P_{j k, j k}(\alpha, \beta)&=p_{j k} f_{j k, j k}(\alpha, \beta), \\
            P_{j k, l m}(\alpha, \beta)&=e^{i \mu_{j k, l m}(t)} f_{j k, l m}(\alpha, \beta),\\
            f_{j k, l m}(\alpha, \beta)&=\delta^{2}\left[\alpha-\alpha_{j k}(t)\right] \delta^{2}\left[\beta-\alpha_{l m}^{*}(t)\right].\\
        \end{aligned}
    \end{equation}
    
    And finally, we substitute Eq. \eqref{smeq:assum_P} into Eq. \eqref{smeq:Pjkjk_dt} and Eq. \eqref{smeq:Pjklm_dt} and obtain Eq. \eqref{smeq:photon_dynamics}.

    ~\\
    \noindent	
    \textcolor{blue}{\textit{$ZZ$ interaction.}}---
    In comparison with the results reported in the studies based on the RIP scheme \cite{Cross2015,Paik2016,Puri2016}, our analysis reveals that Eq. \eqref{smeq:photon_dynamics} introduces an additional coupling term.
    This term supports the long-range $ZZ$ interaction, as well as alters the response of the left resonator to the driving field based on the frequency of the right resonator.
    As a result, any effects exerted by the state of the left qubit on its associated resonator can propagate to the right resonator, subsequently modifying the frequency of the right qubit, and vice versa.
    To quantify the strength of the $ZZ$ interaction in our model, we provide a solution derived from the theoretical framework as delineated by Eq. \eqref{smeq:photon_dynamics}.
    
    First, in this model, we have relations as follows
    \begin{equation}
        \begin{aligned} \label{smeq:EqS13}
        \tilde{\chi}_{00}^{(l)}&=\tilde{\chi}_{01}^{(l)}=\chi^{(l)},\\
        \tilde{\chi}_{00}^{(c)}&=\tilde{\chi}_{01}^{(c)}=\tilde{\chi}_{10}^{(c)}=\tilde{\chi}_{11}^{(c)},\\
        \tilde{\chi}_{00}^{(r)}&=\tilde{\chi}_{10}^{(r)}=\chi^{(r)}\\, 
        \end{aligned}
    \end{equation}
    hence obtaining $ZZ$ interaction strength written as
    \begin{equation} \label{smeq:EqS14}
        \dot{\theta}=\bar{\chi}^{(l)}\left[\alpha_{11}^{(l)}-\alpha_{10}^{(l)}\right] \alpha_{00}^{(l)*}+\bar{\chi}^{(r)}\left[\alpha_{11}^{(r)}-\alpha_{01}^{(r)}\right] \alpha_{00}^{(r)*}
    \end{equation}
    where $\bar{\chi}^{(l)}=\chi_{10}^{(l)}-\chi_{00}^{(l)}$, $\bar{\chi}^{(r)}=\chi_{01}^{(r)}-\chi_{00}^{(r)}$.
    Moreover, the evolution equation for the photon number in Eq. \eqref{smeq:photon_dynamics} can be written as
    \begin{equation} \label{smeq:EqS15}
        \dot{\vec{\alpha}}_{j k}=\hat{G}_{j k} \cdot \vec{\alpha}_{j k}+\vec{E}
    \end{equation}
    where $\vec{\alpha}_{jk}=(\alpha^{(l)}_{jk}, \alpha^{(c)}_{jk}, \alpha^{(r)}_{jk})$, $\vec{E}=(\epsilon^{(l)}, 0, -\epsilon^{(r)})$.
    
    It is assumed, without loss of generality, that the coupling between driving resonators is identical, and that dispersive shifts are equally matched.
    The driving phases of the left and right resonators are set to in-phase and quadrature, a situation referred to as IQ driving.
    This setup promotes photon localization in the driving resonators, thereby minimizing decoherence effects induced by the central resonator.
    As shown in Fig. 4 in main text, the driving phases are pivotal for enhancing the strength of the $ZZ$ interaction.
    
    Next, according to the states of qubits, the corresponding $\hat{G}$ can be obtained
    \begin{equation}
        \begin{aligned}  \label{smeq:EqS16}
            \hat{G}_{00} = &\left[\begin{matrix}\Delta & g & 0 \\ g & \Delta & g \\ 0 & g & \Delta\end{matrix}\right],&
            \hat{G}_{01} = &\left[\begin{matrix}\Delta & g & 0 \\ g & \Delta & g \\ 0 & g & \Delta-\bar{\chi}\end{matrix}\right] \\
            \hat{G}_{10} = &\left[\begin{matrix}\Delta-\bar{\chi} & g & 0 \\ g & \Delta & g \\ 0 & g & \Delta\end{matrix}\right], &
            \hat{G}_{11} = &\left[\begin{matrix}\Delta-\bar{\chi} & g & 0 \\ g & \Delta & g \\ 0 & g & \Delta-\bar{\chi}\end{matrix}\right] \\
        \end{aligned}
    \end{equation}
    
    Finally, when $g \gg \bar{\chi}$, we can obtain at the steady state
    \begin{equation}\label{eq_thetazz}
        \begin{aligned}
        \mathrm{Re}[\dot{\theta}_{zz}] \approx & \frac{\tilde{\epsilon}^2\bar{\chi}^2}{4\Delta}[\frac{9}{9\Delta^2-18g^2-6\Delta\bar{\chi}+\bar{\chi}^2}-\frac{2}{2\Delta^2-3\Delta\bar{\chi}+\bar{\chi}^2}]
        \end{aligned}
    \end{equation}
    \begin{equation}\label{eq_coherence}
        \begin{aligned}
             \mathrm{Im}[\dot{\theta}_{zz}] \approx &\frac{\tilde{\epsilon}^2\bar{\chi}^2\kappa}{4\Delta}[\frac{4\Delta-3\bar{\chi}}{2(\Delta^2-3\Delta\bar{\chi}+\bar{\chi}^2)^2}-\frac{27(3\Delta-\bar{\chi})}{(9\Delta^2-18g^2-6\Delta\bar{\chi}+\bar{\chi})^2}]
         \end{aligned}
    \end{equation}
    Eq. \eqref{eq_thetazz} demonstrates that the $ZZ$ interaction is comparable to that in the previous RIP scheme.
    Specifically, by setting $\epsilon/2\pi = 300$ MHz, $\bar{\chi}/2\pi = 10$ MHz, $\Delta/2\pi = 100$ MHz, $g/2\pi = 100$ MHz, we calculate the $ZZ$ interaction in the model to be $4.74$ MHz, compared to $5.3$ MHz in the previous RIP scheme.
    The $ZZ$ interaction in our scheme is approximately equivalent to that achieved with the prior RIP gate.

\begin{figure*}[htbp]
      \begin{minipage}[b]{1.0\textwidth}
        \centering
        \includegraphics[width=15cm]{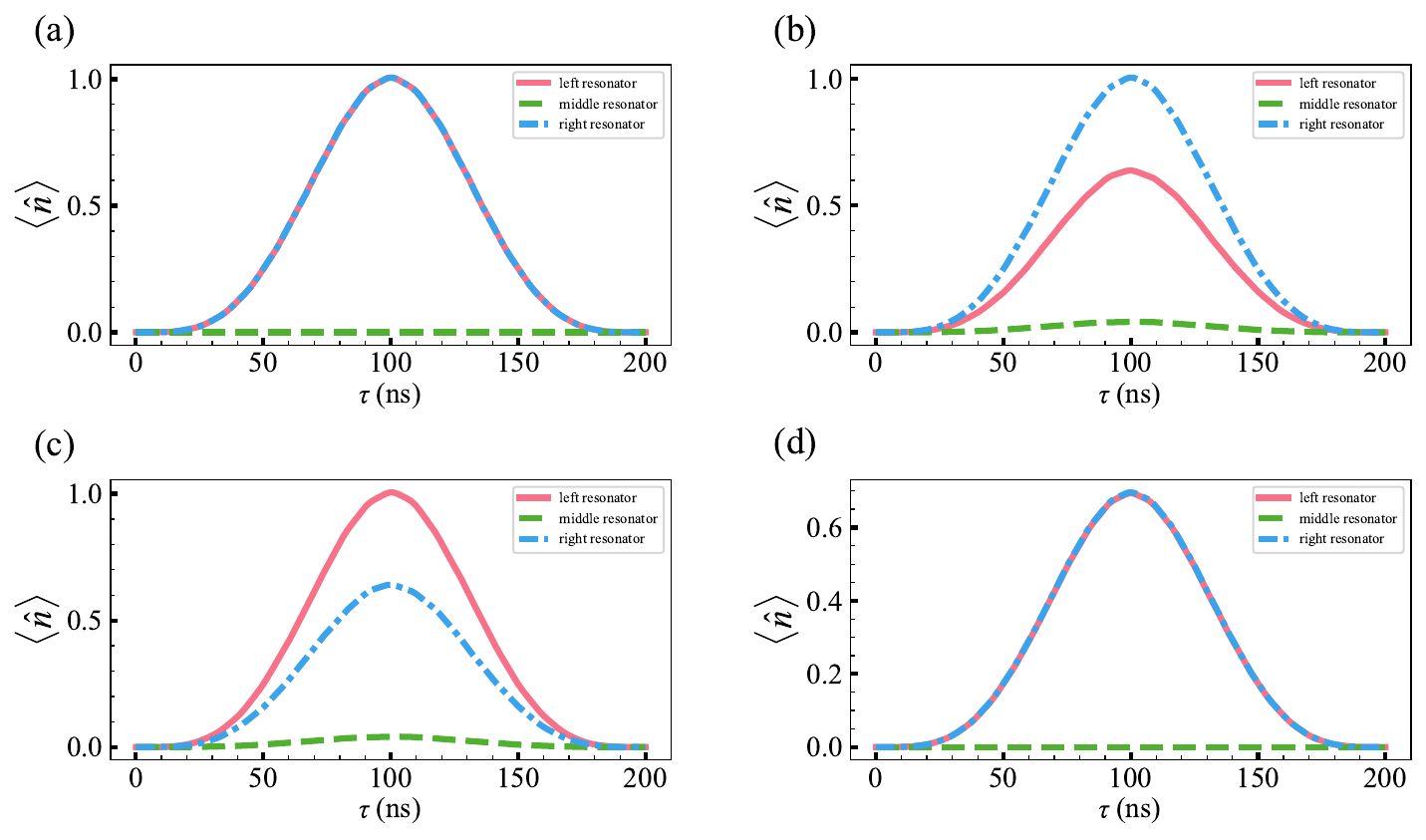}
      \end{minipage}
      \caption{
            (Color online) Evolution of photon numbers in three resonators, with system parameters specified in the first row of Table \ref{tab:sample_params}.
            (a)-(d) depict qubit states $|00\rangle$, $|01\rangle$, $|10\rangle$, and $|11\rangle$, respectively.
            The influence of the central resonator is negligible, demonstrating that photon variation is resilient to its not good performance, including effects like thermal photons and defects causing decoherence.
            \label{smFigure_photon}
        }
    \end{figure*}

    In Fig. \ref{smFigure_photon}, we demonstrate the evolution of photon numbers in three resonators, conditioned on different qubit states.
    The photon distribution is predominantly localized in the driving resonaters, indicating minimal influence from potential losses in the central resonator on the qubit's behavior.
    
    ~\\
    \noindent	
    \textcolor{blue}{\textit{Multi-mode considered.}}---
    When the detuning $|\tilde{\omega}^{(q)} - \nu_m^{(c)}| \gg J_{m}^{(q,c)}$, the leakage of qubit to the low-energy modes can be neglected.
    Then we can found Eq. \eqref{smeq:photon_dynamics} is still valid to the complete model. 
    Slightly different from that in the reduced model, more high-energy modes are involved instead of single mode. 
    This means that the matrix equation in Eq. \eqref{smeq:EqS15} needs to be expanded to higher dimensions to include additional high-energy modes.
    The current question is how many of these additional modes there are.
    By resolving Eq. \eqref{smeq:photon_dynamics}, we found that when the detuning between the long-distance resonator modes and the driving resonator surpasses 13 times their coupling strength, the influence on the ZZ interaction strength between the qubits is negligible.
    Therefore, when the FSR is 13 times greater than $g_m^{(p,c)}$, the reduced model can accurately describe the complete model.
    
    Next, we simply analyzed the leakage of qubit to the low-energy modes.
    Typically, $J_{m}^{(q,c)}/2\pi \approx 10$ MHz, implying a requisite long-distance resonator FSR greater than 200 MHz, corresponding to a physical length of approximately 25 cm.
    To mitigate qubit leakage to the $m$-th mode of the long-distance resonator while extending the long-range coupling, the model incorporates direct coupling between the qubit and the long-distance resonator.
    This direct coupling is notably weaker than that between the qubit and a driving resonator or between the driving and long-distance resonator, hence its negligible influence on $ZZ$ interaction strength.

    \section{Numerical simulations}
    
    \noindent	
    \textcolor{blue}{\textit{Simulation overhead.}}---
    In the reduced model comprising two qubits and three resonators, the dimensions describing the resonator dynamics become substantially large when numerous photons are present.
    We approximate that under steady-state conditions, the resonator microwave photon field adheres to a Poisson distribution.
    Specifically, a resonator with an average photon number around 1 requires modeling with 7 levels, whereas one averaging about 2 photons necessitates 13 levels, to achieve an error below 0.1\%.
    
    Moreover, with the qubits modeled using three levels, the matrix dimensions required for capturing the dynamics accurately swell to 3087$\times$3087 for average photon $\bar{n}=1$, and escalate to 19773$\times$19773 for $\bar{n}=2$.
    Solving the Schr\"{o}dinger equation within such vast dimensional spaces is impractical.
    Consequently, simulations are constrained to an average photon close to 1.
    Simulations reveal that the minimal resonator level descriptions lead to photon leakage into the qubits via the Landau-Zener effect; however, increasing the number of levels mitigates this issue.
    Reducing levels necessitated increasing the qubit-resonator detuning.
    To augment $ZZ$ interaction strength, a large dispersive shift was employed, albeit at the cost of not fully satisfying the dispersive condition.

    \begin{figure*}[htbp]
      \begin{minipage}[b]{1.0\textwidth}
        \centering
        \includegraphics[width=15cm]{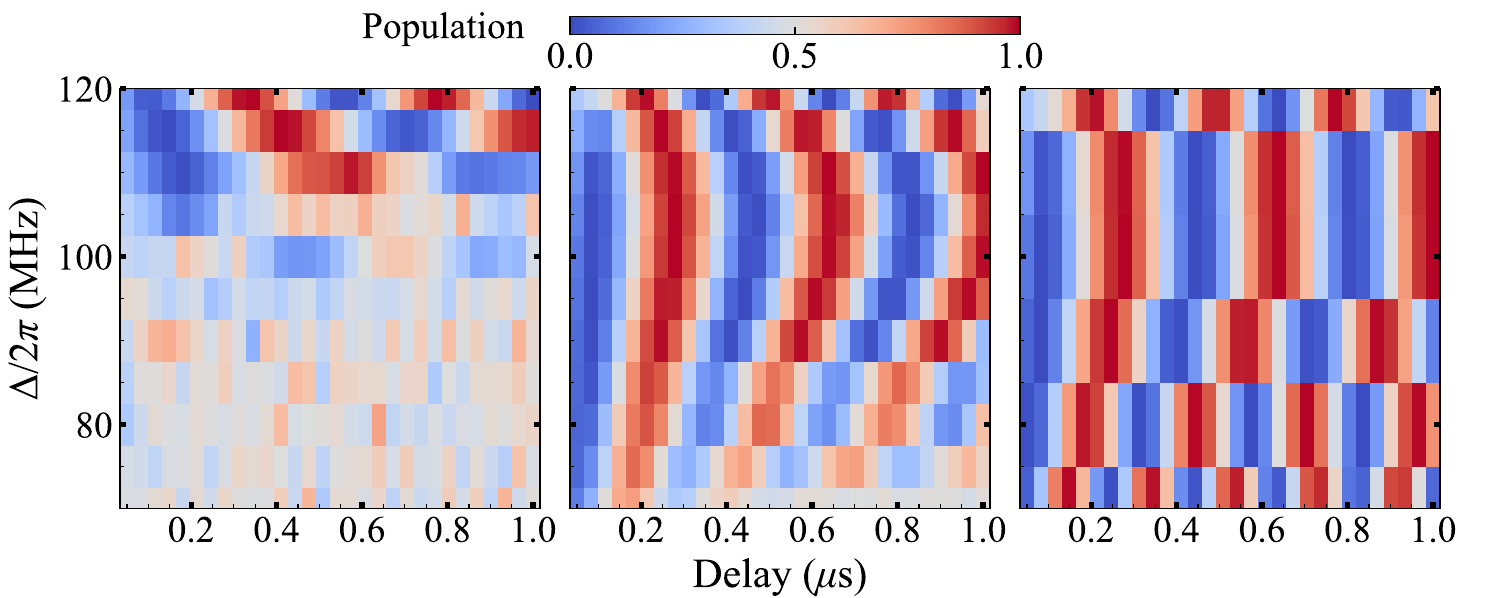}
      \end{minipage}
      \caption{
        (Color online) The evolution of the target qubit population varies with the detuning of the driving field under the resonator with three different energy level truncation numbers (from left to right: 5, 7, 9).
            Using the standard method for controlled phase calibration as elucidate in main text, the population evolution indicate the $ZZ$ interaction. Here, the amplitude of driving fields is $\Omega/2\pi = 100$ MHz. 
        \label{smFigure_evolution}
        }
    \end{figure*}
    
    ~\\
    \noindent	
    \textcolor{blue}{\textit{Supplemental results.}}---
    Based on these considerations, to validate our theoretical model, we adopted the simulation parameters listed in the first row of Table \ref{tab:sample_params}.
    We conducted numerical simulations, delineated by the Hamiltonian in Eq. \ref{smeq:H_disp}.
    To minimize the residual photon within resonators, we adopted the envelope function for the driving field as described by \eqref{smeq:EqS26}, where the degree of the polynomial function sets as $d=3$. 
    
    First, Fig. \ref{smFigure_evolution} validates the numerical simulation adequacy by showcasing the evolution of the controlled phase under RIP sequence timing and driving field detuning for the logn-distance resonator descriptions at level numbers 5, 7, and 9, respectively, from left to right.
    The dynamics of the reduced model are accurately represented when the level number describing the long-distance resonator reaches 9.
    Furthermore, as depicted in Fig. 4 of the main text, the $ZZ$ interaction strength derived from simulations aligns with theoretical predictions, affirming the validity of our theoretical framework, even within the near-dispersive regime.

\begin{figure*}[htbp]
      \begin{minipage}[b]{1.0\textwidth}
        \centering
        \includegraphics[width=15cm]{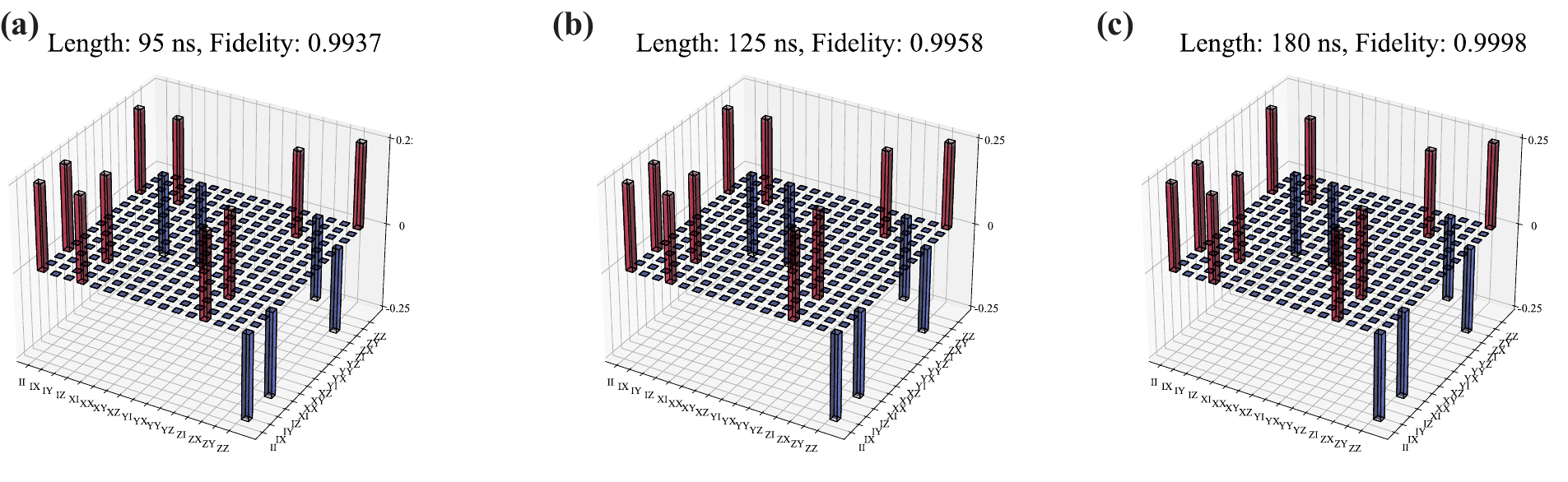}
      \end{minipage}
      \caption{
        (Color online) Reconstructed process matrices of the CZ gate obtained via quantum process tomography in simulation.
                (a) displays results for a CZ gate duration of 95 ns with a fidelity of 99.37\%.
                (b) shows a gate time of 125 ns achieving a fidelity of 99.58\%.
                (c) details a gate time of 180 ns with a fidelity of 99.98\%. 
              \label{smFigure_qpts}
        }
    \end{figure*}

    Furthermore, quantum process tomography (QPT), as shown in Fig. \ref{smFigure_qpts}, was employed to determine the fidelity of the CZ gate.
    Our scheme successfully achieves a CZ gate fidelity of 99.37\% in 100 ns and 99.98\% in 180 ns.

    \section{Optimization of experimental sample parameters}
    
    In this section, we selected various parameters to solve Eq. \eqref{smeq:photon_dynamics}, aiming to satisfy experiment requirements.
    Firstly, residual photons induce additional qubit decoherence in the subsequent gate operation.
    To mitigate this, an optimal drive field envelope function and frequency were selected to minimize residual photon numbers.
    
    \begin{figure*}[htbp]
      \begin{minipage}[b]{1.0\textwidth}
        \centering
        \includegraphics[width=15cm]{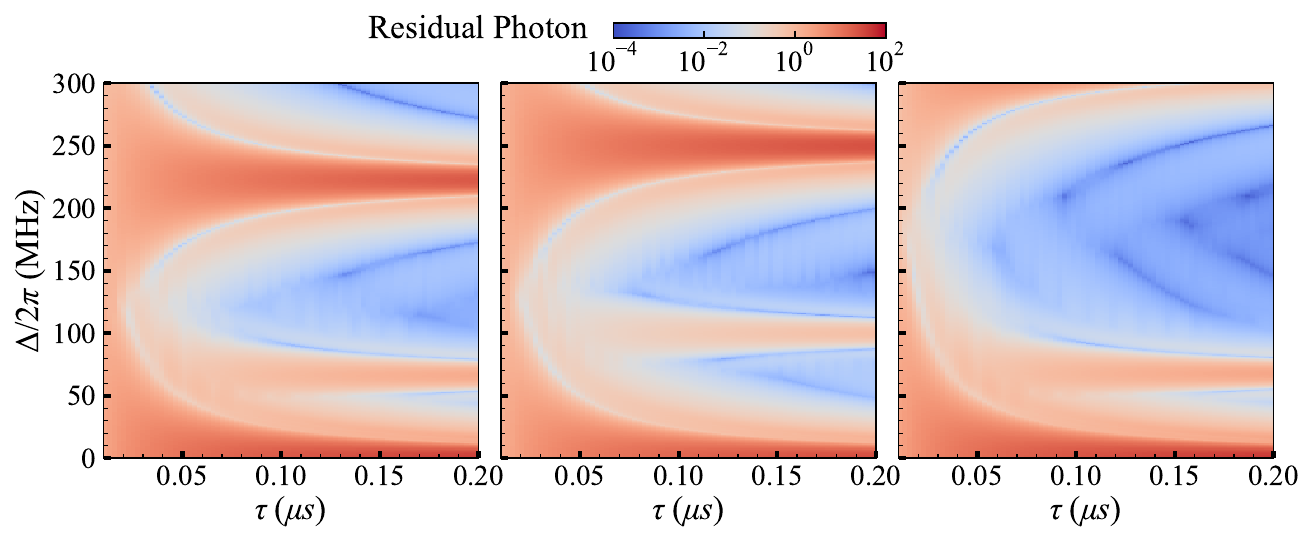}
      \end{minipage}
      \caption{
            (Color online) Variation of residual photon numbers as a function of evolution time and detuning of the driving field.
                (a) corresponds to a free spectral range of 200 MHz and a coupling strength $g^{r,c}/2\pi = 40$ MHz.
                (b) is for an FSR of 200 MHz with $g^{r,c}/2\pi = 80$ MHz.
                (c) shows results for an FSR of 300 MHz and $g^{r,c}/2\pi = 40$ MHz.
            \label{smFigure_rps}
        }
    \end{figure*}
    
    We first explored the photon number evolution by reformulating Eq. \eqref{smeq:photon_dynamics} into Eq. \eqref{smeq:EqS15}.
    This involved employing the unitary transformation $\hat{U}_{jk}$ to diagonalize the matrix $\hat{G}_{jk}$, leading to the derivation
    \begin{equation}
        \begin{aligned} \label{smeq:EqS24}
            \dot{\vec{\beta}}_{jk}=\hat{D}_{jk}\cdot\vec{\beta}_{jk}+\hat{U}_{jk}\cdot\vec{E}
        \end{aligned}
    \end{equation}
    Here, $\hat{D}_{jk}=\hat{U}_{jk}\hat{G}_{jk}\hat{U}_{jk}^{\dagger}$ represents the diagonal matrix with elements
    $\tilde{\Delta}_{jk}^{(p)} = i\left(\Delta_{d}^{p}+\tilde{\chi}_{j k}^{(p)}\right)+\kappa^{(p)}/2$,
    where $\Delta_{d}^{p}$ denotes the detuning between the drive field frequency and the decoupled resonator frequency for mode $p$.
    Eq. \eqref{smeq:EqS24} captures the evolution dynamics of each decoupled resonator in its eigenbasis,
    facilitating the calculation of photon number evolution for any given resonator mode $p$, and we obtain
    \begin{equation}\begin{aligned} \label{smeq:EqS25}
        \beta_{jk}^{(p)}(t)=-\frac{i}{2}e^{-\tilde{\Delta}_{jk} t}\int_0^{t}e^{\tilde{\Delta}_{jk} t'}\tilde{\varepsilon}_{jk}^{(p)}(t') dt'
    \end{aligned}\end{equation}
    where $\tilde{\varepsilon}^{(p)}(t) = \{\hat{U}_{jk}\cdot\vec{E}\}_{p}$ is the drive field applied to the decoupled resonator $p$.

    \begin{figure*}[htbp]
      \begin{minipage}[b]{1.0\textwidth}
        \centering
        \includegraphics[width=10cm]{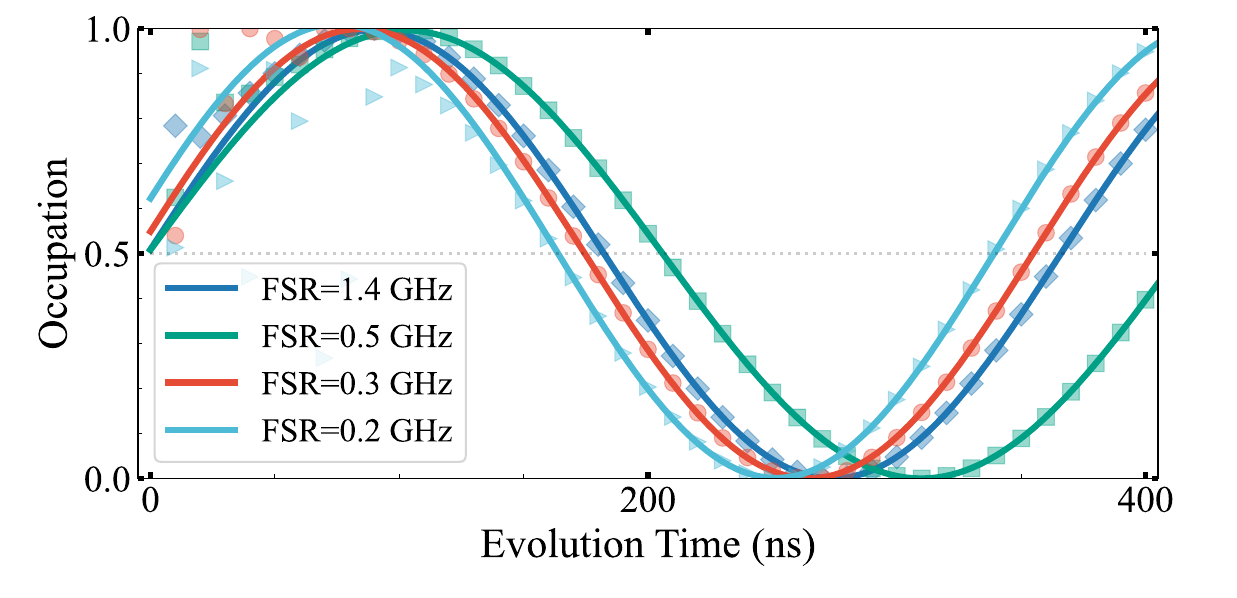}
      \end{minipage}
      \caption{
            (Color online) Evolution of the target qubit, using the sample parameters specified in Table \ref{tab:sample_params}.
            \label{smFigure_evolution_vs_fsr}
        }
    \end{figure*}
    
    \begin{table}[htbp]
      \centering
        \caption{Designed sample parameters and suggested driving parameters achieve FSR and CZ gates within approximately 200 ns.
                For brevity, variables $\nu^{l,r}$,  $\omega^{l,r}$, $\eta$, $g^{p,q}$, $g^{p,c}$, $\Delta$, $\epsilon$ denote frequency, not angular frequency.}
      \label{tab:sample_params}
      \begin{tabular}{cc|c|c|c|c|c|c|cc}
            \hline\hline
            & FSR (GHz)& $\nu^{l}, \nu^{r}$ (GHz) & $\omega^{l}, \omega^{r}$ (GHz) & $\eta$ (MHz) & $g^{ll}, g^{rr}, g^{(c,p)}$ (MHz)  & degree $d$ & $\Delta, \epsilon$ (MHz) & CZ gate time (ns) \\ \hline
            & 1.40     &6.9239, 6.9168     & 5.000, 4.800         & -280           & 390, 428, 100          & 3 & 100,                   200 & 180         \\
            & 0.50     &5.9808, 5.9820     & 5.250, 5.200         & -280           & 120, 120, 80           & 3 & 70,                    280 & 205          \\
            & 0.30     &5.9796, 5.9795     & 5.150, 5.155         & -280           & 130, 130, 60           & 7 & 50,                    200 & 175          \\
            & 0.20     &5.9854, 5.9853     & 5.300, 5.305         & -280          & 100, 100, 50            & 9 & 40,                    200 & 165          \\
            \hline\hline
      \end{tabular}
    \end{table}

    Next, we employed the drive field envelope delineated by Eq. \eqref{smeq:EqS26}, where $d=3$.
    This situation elucidated the relationship between the residual photon number, the gate operation time, and the drive field detuning $\Delta_d$, as illustrated in Fig. \ref{smFigure_rps}. And other parameters are shown in row 4 of Table. \ref{tab:sample_params}.
    Analysis revealed that lower residual photon regions are segmented by the frequencies of the dressed resonators into distinct intervals.
    By reducing FSR of the long-distance resonator, the frequency bands associated with transitions (red bands) condense, enhancing the spectral resolution.
    Increasing the coupling strength between the resonators subsequently expands the detuning between the zeroth and first red bands.
    This broader detuning interval, deemed the operational interval, is critical for effective $ZZ$ interaction, underscoring the need for an expansive frequency range within this interval.

\begin{figure*}[htbp]
      \begin{minipage}[b]{1.0\textwidth}
        \centering
        \includegraphics[width=15cm]{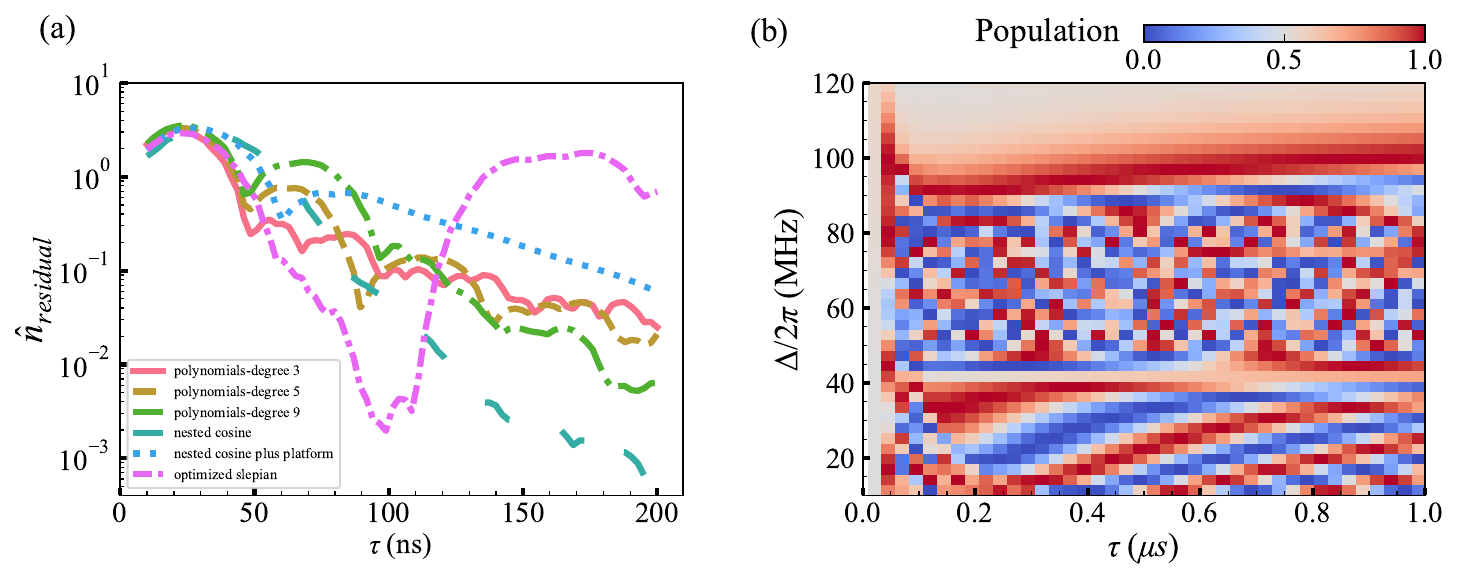}
      \end{minipage}
      \caption{
        (Color online)
            (a) Residual photon numbers as a function of evolution time for various envelope functions, including polynomials with $d=3,5,9$, nested cosine, and optimized Slepian, displayed with different colored lines.
                Notably, optimization of driving pulses can achieve a fast CZ gate within 100 ns with residual photon numbers approaching $10^{-3}$.
            (b) Evolution of the target qubit population as a function of driving field detuning during optimization of the detuning parameter.
                For detunings between 10 MHz and 45 MHz, the target qubit exhibits rapid, coherent oscillations, indicating a good performance controlled phase gate can be achieved.
              \label{smFigure_others}
        }
    \end{figure*}

    To optimize this scenario, parameters from row 4 of Table. \ref{tab:sample_params} were adopted alongside varied drive field envelope functions aimed at minimizing the residual photon count.
    The results, as shown in Fig. \ref{smFigure_others}(a), indicate that a third-degree polynomial waveform achieves the lowest residual photon count swiftly, whereas a ninth-degree polynomial provides a lower count over extended durations.
    Nonetheless, these methods did not reduce the residual photon below 1\% in the desired timeframe.
    Following the fast adiabatic pulse proposed by Martinis and Geller in 2014 \cite{Martinis2014}, differing with the method used in prior RIP works \cite{Cross2015,Paik2016,Puri2016}, we implemented Eq. \eqref{smeq:EqS26} to design an optimized waveform that successfully suppressed the residual photon to below 1\% within approximately 100 ns, as discussion in the following section.
    
    And last, we study the impact of drive field detuning on $ZZ$ interaction strength, employing the sample parameters from row 4 of Table \ref{tab:sample_params}, as depicted in Fig. \ref{smFigure_others}(b).
    The analysis identified distinct behaviors of the target qubit across three detuning ranges within the driving field detuning $\Delta/2\pi$.
    For detunings between 10 MHz and 45 MHz, the target qubit demonstrated rapid, coherent oscillations.
    Conversely, within 45 MHz to 90 MHz, these oscillations became erratic, and for detunings between 90 MHz and 120 MHz, the qubit exhibited minimal oscillations.
    
    The optimal region, where coherent oscillations occur, aligns with the dressed state resonator frequency.
    This region corresponds to the frequency gap between the zeroth and first red bands, as outlined earlier in Fig. \ref{smFigure_rps}.
    This specific detuning interval is crucial for the effective implementation of our long-range, high-fidelity CZ gate.

    \section{Optimization of driving pulse parameters}

    We analyzed the driving field envelope function, considering various structured forms to optimize pulse shaping within quantum computing applications. The functional forms include piecewise polynomials, nested cosine shapes, nested cosine with a platform, and optimized Slepian shapes, each tailored for specific operational conditions.
    First, the piecewise polynomial shapes are given by
    \begin{equation}\begin{aligned} \label{smeq:EqS26}
         \epsilon(t)=\sum_{m=0}^{(d-1)/2}(-1)^mc_m(t/t_r)^{m+(d+1)/2}\\
         \sum_{m=0}^{(d-1)/2}(-1)^mc_m\binom{(d+1)/2+m}{j}j!=\delta(j)
    \end{aligned}\end{equation}
    where $\delta(j)$ takes the value 1 if $j=0$, and 0 otherwise.
    The order of the piecewise polynomial function is indicated by the degree $d$, and $2t_r$ is the pulse duration.
    
    Next, nested cosine shapes are described by
    \begin{equation}\begin{aligned} \label{smeq:EqS27}
         \epsilon(t)=(1+\cos[\pi \cos(\frac{\pi  t}{2t_r})])/2
    \end{aligned}\end{equation}
    where $2t_r$ is the pulse duration.
    
    Then, combined the nested consine and platform shapes, the pulse can be cast as
    \begin{equation} \label{smeq:EqS28}
    \epsilon(t) = 
    \begin{cases}
        (1+\cos[\pi \cos(\frac{\pi  t}{2t_r})])/2 & \text{if } 0\leq t < t_r \\
        1 & \text{if } t_r\leq t < t_r+t_p \\
        (1+\cos[\pi \cos(\frac{\pi  (t-t_p)}{2t_r})])/2 & \text{if } t_r+t_p\leq t \leq 2t_r+t_p 
    \end{cases}
    \end{equation}
    where $2t_r+t_p$ is the pulse duration and $t_p/t_r=0.2$ in Fig. \ref{smFigure_others}(a).

    \begin{figure*}[htbp]
      \begin{minipage}[b]{1.0\textwidth}
        \centering
        \includegraphics[width=8.5cm]{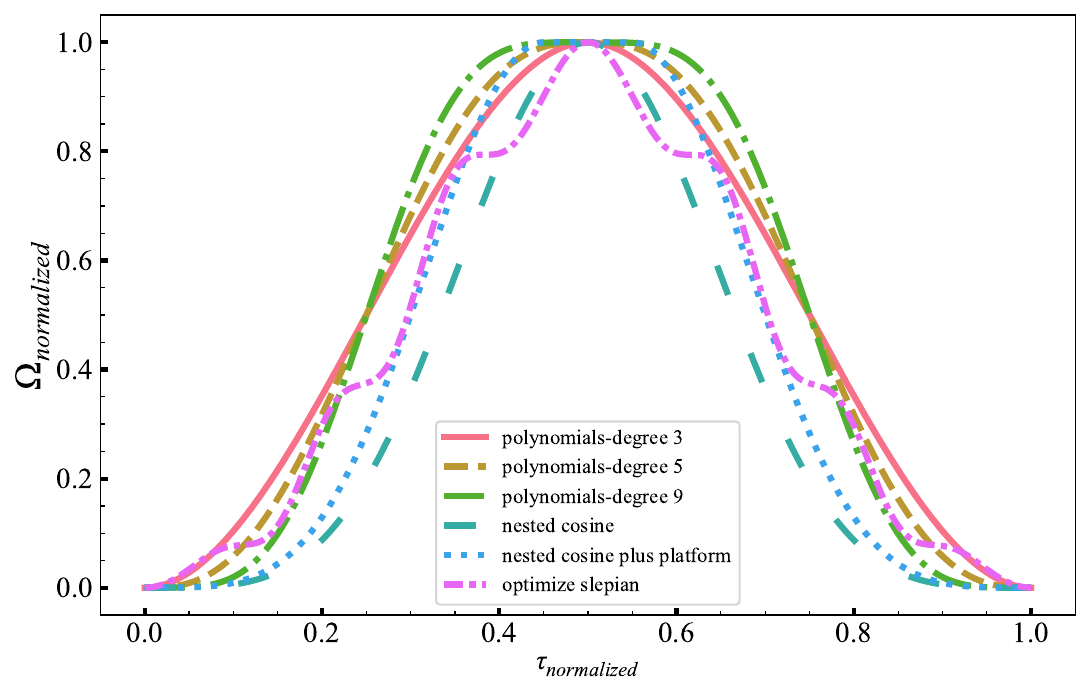}
      \end{minipage}
      \caption{
        (Color online)
            Normalized amplitudes of driving pulses under various envelope functions, resulting the residual photon numbers depicted in Fig. \ref{smFigure_others} for achieving high-performance CZ gates across different operational speeds.
            \label{smFigure_pulse}
        }
    \end{figure*}
    
    Lastly, the optimized Slepian shapes are described by
    \begin{equation}\begin{aligned} \label{smeq:EqS29}
         &\epsilon(t)=\sum_{j=0}^{n} \lambda_j\{1-\cos[\pi(j+1)t/t_r]\}/2 \\
         &\sum_{j\in odd} \lambda_j = 1
    \end{aligned}\end{equation}
    where coefficients $\lambda$ are optimized using a differential evolution algorithm for minimal residual photon presence (below 1\%),
    aiming for gate times around 100 ns, as shown in Fig. \ref{smFigure_others}(a), with coefficients $\lambda$=[0.9429, -0.089, -0.003, 0.0002, -0.0003, -0.0364, 0.0835].
    In addition, these envelope functions, depicted in Fig. \ref{smFigure_pulse} correspond to the results shown in Fig. \ref{smFigure_others}(a).


\end{document}